\pdfoutput=1
\documentclass[a4paper,10pt]{article}

\usepackage{fullpage} 
\usepackage{parskip} 
\usepackage{tikz} 
\usepackage{amsmath}
\usepackage{hyperref}
\usepackage{graphicx}
\usepackage[font=footnotesize]{caption}
\usepackage{array}
\usepackage{multicol}
\usepackage{amssymb}
\usepackage[ruled,vlined,linesnumbered]{algorithm2e}
\usepackage{mathrsfs}
\usepackage{bm}
\SetAlFnt{\footnotesize}
\usepackage{adjustbox}
\graphicspath{ {images/} }
\usepackage{color}
\usepackage{hyperref}
\usepackage{booktabs}
\usepackage{multirow}
\usepackage[numbers]{natbib}

\usepackage{subfiles}

\newcommand{\argmin}{\operatornamewithlimits{arg\ min}}

\definecolor{darkgreen}{rgb}{0,0.6,0.2}
\definecolor{orange}{rgb}{1.0,0.5,0.0}

\DeclareCaptionFormat{myformat}{\selectfont#1#2#3}
\title{Segment-Based Credit Scoring \\ Using \\ Latent Clusters in the Variational Autoencoder}
\author{Rogelio A. Mancisidor$^{a,b,\ast}$, Michael Kampffmeyer $^a$, 
Kjersti Aas $^c$, Robert Jenssen $^a$\\
\\
\normalsize{$^{a}$UiT Machine Learning Group, Faculty of Science and Technology,}\\
\normalsize{Department of Physics and Technology, University of Troms{\o}, 9019 Troms{\o} Norway,}\\
\normalsize{$^b$Credit Risk Models, Santander Consumer Bank AS, 1325 Lysaker Norway}\\
\normalsize{$^{c}$Statistical Analysis, Machine Learning and Image Analysis,}\\
\normalsize{Norwegian Computing Center, 0373 Oslo Norway}\\
\\
\normalsize{$^\ast$Corresponding author; E-mail: rogelio.a.mancisidor@uit.no}
\
}

\begin{document}

\usetikzlibrary{decorations.pathreplacing}
\usetikzlibrary{fadings}
\usetikzlibrary{arrows.meta, bending, backgrounds, scopes}
\usetikzlibrary{positioning, fit, arrows.meta}
\def\layersep{1.8cm}

\maketitle

\renewenvironment{abstract}
{\begin{quote}
\noindent \rule{\linewidth}{.5pt}\par{\bfseries \abstractname.}}
{\medskip \noindent 
\rule{\linewidth}{.5pt}
\end{quote}
}

\begin{abstract}
Identifying customer segments in retail banking portfolios with different risk profiles can improve the accuracy of credit scoring.  The Variational Autoencoder (VAE) has shown promising results in different research domains, and it has been documented the powerful information embedded in the latent space of the VAE. We use the VAE and show that transforming the input data into a meaningful representation, it is possible to steer configurations in the latent space of the VAE. Specifically, the Weight of Evidence (WoE) transformation encapsulates the propensity to fall into financial distress and the latent space in the VAE preserves this characteristic in a well-defined clustering structure. These clusters have considerably different risk profiles and therefore are suitable not only for credit scoring but also for marketing and customer purposes. This new clustering methodology offers solutions to some of the challenges in the existing clustering algorithms, e.g., suggests the number of clusters, assigns cluster labels to new customers, enables cluster visualization, scales to large datasets, captures non-linear relationships among others. Finally, for portfolios with a large number of customers in each cluster, developing one classifier model per cluster can improve the credit scoring assessment. 

\end{abstract}

\section{Introduction}\label{Intro}
Lending is the principal driver of bank revenues in retail banking, where banks must assess whether to grant a loan at the moment of application. Therefore, banks focus on different aspects to improve this credit assessment. Understanding customers and target populations can improve risk assessment \cite{siddiqi2012credit}. For example, bank analysts possess the knowledge to understand the underlying risk drivers. This information can be used to identify different groups within a given portfolio, i.e. heuristic-based segmentation \cite{siddiqi2012credit}, and carry out the risk assessment segment-wise.

However, identifying different segments is not an easy task, and in case any segmentation has been chosen using the heuristic-based approach, this segmentation does not guarantee better risk assessment \cite{makuch2001basics}. Hence, analysts need quantitative tools that can help them to identify groups with distinct risk profiles and not only different demographic factors. The presumption is that these groups have different levels of creditworthiness and the factors underlying their risk profiles are different \cite{siddiqi2012credit}. Hence, segment-based credit scoring in some cases can increase the overall model performance compared to a single portfolio-based model.

Different quantitative techniques can be used to identify distinct segments. Principal component analysis (PCA) can find clustering structures by applying orthogonal linear transformations. Hence, PCA does not necessarily model complex non-linear relationships \cite{van2009dimensionality}. K-means \cite{lloyd1982least} is another popular methodology which aims to represent data by $k$ centroids. Choosing the optimal $k$ value is challenging in practice, and it does not offer the possibility to visualize the clustering structure. The existing clustering methodologies have some limitations which are unappealing in the segment-based scoring approach, see for example \cite{ilango2011cluster,agarwal2011issues,salvador2004determining} or Section (\ref{clustering_analysis})

The Variational Autoencoder (VAE) \cite{kingma2013auto,rezende2014stochastic} has shown promising results in different research domains. In some cases it outperforms state-of-the-art methodologies, e.g., in the medical field predicting drug response \cite{rampasek2017dr} or classifying disease subtypes of DNA methylation \cite{bioinformatics18}, in speech emotion recognition \cite{latif2017variational}, in generative modeling for language sentences imputing missing words in large corpus \cite{bowman2015generating}, in facial attribute prediction \cite{hou2017deep} among others. Interestingly, all this research documents the powerful information embedded in the latent space of the VAE.  

We use the VAE and the Auto Encoding Variational Bayesian (AEVB) algorithm \cite{kingma2013auto} in this research to identify hidden segments in customer portfolios in the bank industry. Given that the VAE has converged to the optimal variational density, the evidence lower bound (ELBO) (See Section \ref{sec_vi}) generates codes in the latent space which are likely to generate the input data. Otherwise, the algorithm would not have converged in the first place. Hence, by transforming the input data into a meaningful representation, we can indirectly steer configurations in the latent space. The main contribution of this research is a segmentation approach which offers an effective manifold learning algorithm \cite{Goodfellow-et-al-2016} and captures valuable information into the latent space of the VAE. Hence, this technique is well suited for analyzing high-dimensional financial data with complex non-linear relationships \cite{khandani2010consumer}. Moreover, this methodology provides solutions to some of the problems in the existing clustering algorithms, e.g., suggests the number of clusters, assigns cluster labels to new customers, enables cluster visualization, scales to large datasets and does not require expert domain to choose input parameters. Banks can use these clusters for marketing and customer purposes \cite{anderson2007credit} and, given that the clusters have a considerably different risk profile, they can also be used for segment-based credit scoring. 

This paper is organized as follows. Section \ref{sec_literature} reviews the related work in credit risk segmentation, while Section \ref{sec_ae} explains the main characteristics of feature learning. Section \ref{sec_vae} introduces Variational Inference and shows the derivation of the ELBO together with its properties. Section \ref{sec_segmentation} explains the data transformation used to steer configurations in the latent space and Section \ref{sec_results} presents the experiments conducted and findings. Finally, Section \ref{sec_conclusion} presents the main conclusions in this paper.

\section{Related Work}\label{sec_literature}
There exist five factors that could bring up the need for different segments i) Marketing, ii) Customer, iii) Data, iv) Process and v) Model fit \cite{anderson2007credit}. Marketing factors arise where banks require greater confidence in a specific segment to ensure ongoing health of the business. Customer factors apply where banks want to treat customers with particular characteristics separately, e.g., customers with no delinquency records. The data factors relate to different operating channels, e.g., internet or dealer, where application data can be entirely different, and process factors refer to the fact that there are products that are treated differently from a managerial point of view. Finally, model fit factors are interactions within the data where different characteristics predict differently for different groups. In this research, we focus on model fit factors.

Segmentation can be done using a heuristic or statistical-based approach \cite{siddiqi2012credit}. In the heuristic-based approach, bank analysts define segments based on both customer characteristics and using expert knowledge of the business. For example, a bank can segment customers based on whether they have previous delinquency records. Further, based on these two segments it is the responsibility of a classification model to separate \textit{good} from \textit{bad} customers. When classification is the final goal of segmentation, segments should not only be identified based on demographic factors, but rather on risk-based performance \cite{siddiqi2012credit}. Following this line of thought, \cite{makuch2001basics} posit that heuristic-based segmentation does not necessarily improve the classification results.

One example of the heuristic-based approach is the dual scoring model developed in \cite{chi2012hybrid}.  They divide customers based on whether they have previous delinquency marks. Then, they use one logistic regression model in each segment to create a credit score. This score is further combined with a behaviour score to create a risk matrix with these two dimensions. Note that the segmentation, in this case, is based on expert-knowledge and the classification, in a second step, is done based on these segments.

Statistical-based techniques find segments using statistical, data-mining or machine learning methods. In its most basic form, one customer attribute is used to create different customer groups with distinct characteristics. It is also possible to use several variables, either one at a time or multiple features in a multi-dimensional vector, to create such groups. By doing this, it is feasible to achieve a deeper segmentation granularity. 

In \cite{VantageScore} they argue that the heuristic-based approach is sub-optimal because it is not possible to approximate the target population for a given financial institution by merely using one characteristic to create customer segments. Instead, the authors propose a hybrid model where the heuristic-based approach is used together with credit scores estimated using a statistical model. Furthermore, this ranking possesses information which makes it possible to group individuals with similar behaviour along different dimensions. Note that in this approach the final goal is to rank customers based on the segments found by the hybrid method. Hence it is an associative or descriptive approach, not a predictive approach \cite{aurifeille2000bio}.

There are two ways to deal with segment-based credit scoring regardless of segmentation technique, the two-step method and the simultaneous approach. In the two-step method, we find segments in the first step. Then, a classification model, one per group, assesses credit risk. On the other hand, the simultaneous approach optimizes both the segments and the classification task in one single step. Mathematically, the simultaneous approach optimizes only one objective function which takes into account segmentation and classification.

One example of the simultaneous approach is \cite{bijak2012does}, where they use the Logistic Trees with Unbiased Selection (LOTUS) model. The model optimizes both segmentation and classification in a simultaneous step. In one hand, the model identifies interactions within the data to create segments and, on the other hand, this segmentation aims to improve classification performance. The study does not show any significant improvement in model performance when segmentation is considered. 

In a different approach to segmentation, \cite{hand2005optimal} finds optimal bipartite segments which optimize classification performance. The authors apply an exhaustive search on every single possible split point for all variables at hand. These points create the bipartite clusters on which two logistic regression models are built using all variables in the dataset. The optimal cut-off point is the one which maximizes the overall likelihood of the two logistic regression models. In their results, they find significant accuracy improvements, after correcting for class imbalance, using the segment-based approach. Note that the segmentation, in this case, is based on only one customer characteristic. Hence, exhaustive search on each feature in the dataset may become unfeasible for many real datasets.

\section{Feature Learning - Autoencoders}\label{sec_ae}
In the rest of the paper we use the following notation. We consider i.i.d. data $\{x_i\}_{i=1}^n$ where $x_i \in \mathbb{R}^{d_x}$. Further, latent variables $\{z_i\}_{i=1}^n$ where $z_i \in \mathbb{R}^{d_z}$ can be i.i.d. data or simply the result of a deterministic function, this is clear from the context. Finally, we drop the subscript $i$ whenever the context allows it. 

\begin{figure}
    \centering
    {
    \begin{tikzpicture}
    [shorten >=1pt,->,draw=black!50, 
    node distance = \layersep,
every pin edge/.style = {<-,shorten <=1pt},
		rectangle/.style={fill=black!25,minimum size=17pt,inner sep=0pt},
        neuron/.style = {circle,fill=black!25,minimum size=17pt,inner sep=0pt},
  input neuron/.style = {neuron, fill=black,opacity=0.2,draw=black},
 output neuron/.style = {neuron, fill=black,opacity=0.2,draw=black},
 hidden neuron/.style = {neuron, fill=black,opacity=0.2,draw=black},
         annot/.style = {text width=3em, text centered},
                         ]
\foreach \name / \y in {1,...,3}
    \node[input neuron, pin=left:$x_\y$] (I-\name) at (0,-\y) {};

\foreach \name / \y in {1,...,2}
	\path[yshift=-0.5cm]
     	node[hidden neuron] (H-\name) at (\layersep,-\y cm) {};

\foreach \name / \y in {1,...,3}
	\node[output neuron,pin={[pin edge={->}]right:$\tilde{x}_\y$}] (O-\name)  at (2*\layersep, -\y)  {};

\foreach \source in {1,...,3}
    \foreach \dest in {1,...,2}
        \path (I-\source) edge (H-\dest);

\foreach \source in {1,...,2}
	\path (H-\source) edge (O-1);

\foreach \source in {1,...,2}
	\path (H-\source) edge (O-2);

\foreach \source in {1,...,2}
	\path (H-\source) edge (O-3);
	
\draw[-,shorten >= -0.2pt,decorate,decoration={brace,mirror,raise=5pt}]
(0.2,-3.2) --node[below=7pt] {$f_{\phi}(x)$} (1.6,-3.2);

\draw[-,shorten >= -0.2pt,decoration={brace,mirror,raise=5pt},decorate]
(2.1,-3.2) --node[below=7pt] {$g_{\theta}(z)$} (3.3,-3.2);
    
\node[annot,above of=H-1, node distance=1.3cm] (hl) {\small Hidden layer};
\node[annot,left of=hl] {\small Input layer};
\node[annot,right of=hl] {\small Output layer};

    \end{tikzpicture}
}
    \caption{Basic structure for an Autoencoder (AE) which is a feed forward network trained to learn the input data $x$. The parametric function $f_{\phi}(x)$ transforms the input data $x$ into a new representation $z$. Then the parametric function $g_{\theta}(z)$ convert the codes $z$ back to their original space, generating the reconstruction $\tilde{x}$.}
    \label{ae_fig}
\end{figure}
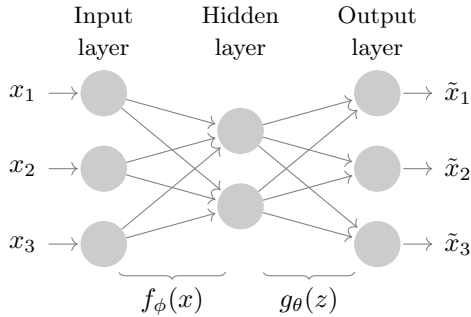

An autoencoder (AE) is a feedforward neural network which is trained to reconstruct the input data, see Figure (\ref{ae_fig}). This is achieved by first using a function $f_{\phi}(x)$ which encodes the input data to a new representation $z$, which is further decoded by $g_{\theta}(z)=\tilde{x}$ into its original space. We say that in this last step the original feature vector $x$ is reconstructed. 

The parameters $\phi,\theta$ in the AE are trained by minimizing the squared reconstruction error, where the reconstruction error is defined as $\epsilon = x-\tilde{x}$. The most basic architecture for an AE is feedforward networks with one input layer, one output layer, and one or multiple hidden layers with non-linear activation functions \cite{rumelhart1986learning}. In the case where linear activation functions are chosen, it has been shown that the squared reconstruction error has a unique local and global minimum. Further, the global minimum is described by principal component analysis and least squares regression. Therefore it can be solved by other algorithms rather than gradient descent and backpropagation \cite{baldi1989neural}.

To have a better understanding of the role of feedforward networks in the AE, Figure (\ref{ae_fig}) shows a basic architecture with only one hidden layer. Between the input layer and the hidden layer is where the encode function $f_{\phi}(x)$ transforms the input data into its new representation $z$ by using the activation function $s_f(b+Wx)$. Further, between the hidden layer and the output layer is where the decoder $g_{\theta}(z)$ generates the reconstruction $\tilde{x}$ using the activation function $s_g(d+W'z)$ \cite{bengio2013representation}. In this case, $W$ and $W'$ are the weights of the encoder and decoder respectively. In the same way, $b$ and $d$ are the bias term of the encoder and decoder. Therefore, $\phi=\{W,b\}$ and $\theta=\{W',d\}$. Note that it is also possible to stack multiple hidden layers in both the encoder and decoder part of the AE. 

The specific architecture in Figure (\ref{ae_fig}) shows a bottleneck architecture. This is probably the simplest approach preventing the AE from learning the identity function, i.e. $x=\tilde{x}$ \cite{bengio2013representation}. However, regularized autoencoders, e.g., denoising, sparse or contractive autoencoders, try to solve this problem using different approaches and not simply using a bottleneck architecture.

The denoising AE corrupts the input data $x$ by means of a stochastic mapping $\tilde{x}\sim q_\mathcal{D}(\tilde{x}|x)$. This corrupted data $\tilde{x}$ is used to train the AE. The main goal is that the network learns the underlying structure of the data generating process $p(x)$ and will not focus on details on the training dataset, hence it learns \textit{good representations}. Note that denoising AE still backpropagates the reconstruction error, but the difference is that a deterministic mapping to a corrupted input obtains the codes. For more details please refer to \cite{vincent2010stacked}.

The contractive autoencoders attenuate the sensitivity of the learned features penalizing the objective function by the Frobenius norm of the encoder's Jacobian. In other words, the contractive AE uses an efficiently computable expression to penalize the objective function, with a fine-tuning hyperparameter, and not a stochastic scheme as in the denoising AE \cite{bengio2013representation}.  

On the other hand, the sparse AE forces the encoder weights to be equal to the decoder weights to restrict capacity. Further, sparsity regularization penalizes the bias terms in the hidden layer or their outputs. Penalizing the bias can come at the cost of hurting the numerical optimization. However, there are different options to penalize the output of the hidden units, e.g. L1 penalty, Student-t penalty, or the average output of hidden units \cite{bengio2013representation}.

Finally, different sort of autoencoders have been used in multiple tasks such as transfer learning \cite{zhuang2015supervised,deng2013sparse}, generating new data \cite{zhang2015learning}, clustering \cite{jiang2017variational,zhang2017deep}, classification tasks based on reconstruction errors \cite{an2015variational}, and for dimensionality reduction.

All of the previous versions of AEs, that we just have mentioned, use feedforward networks to generate latent variables and reconstruct the input data. In this paper we use an extension of the Variational Autoencoder (VAE), which has a probabilistic background. Specifically, the encoder and decoder can take the form of well know probability densities, e.g., Gaussian density functions. Most importantly, the VAE offers an excellent manifold learning algorithm by simultaneously training the probabilistic encoder together with the decoder. This makes the model learn a predictable coordinate system which is encapsulated in the encoder \cite{Goodfellow-et-al-2016}. This property makes the VAE well suited for analyzing high-dimensional credit risk data where complex and non-linear relationships exist \cite{khandani2010consumer}. The VAE is also a generative model, hence we can draw synthetic data by simply passing Gaussian noise to the decoder. This feature is appealing in domains where large amount of data is not available or is missing. 

\section{Variational Autoencoder}\label{sec_vae}
\subsection{Variational Inference}\label{sec_vi}
Consider the joint density $p(x,z)$ where $x$ is the observed data and $z$ is a latent variable. Then in the Bayesian framework, the latent variable $z$ helps govern the data distribution \cite{vi_review}. To understand this, think about the error term $\epsilon$ in the linear model $y=x^T\beta+\epsilon$. The error term is not observed, hence it is latent, but we use it to say something about the dispersion in the data distribution of $y$. 

The latent variable in the joint density $p(x,z)$ is drawn from a prior density $p(z)$ and then it is linked to the observed data through the likelihood $p(x|z)$. Inference amounts to conditioning on data and computing the posterior $p(z|x)$ \cite{vi_review}.

The problem is that this posterior is intractable in most cases. Note that
\begin{equation}
p(z|x) = \frac{p(z,x)}{p(x)},
\label{true_post}
\end{equation}
where the marginal distribution is $p(x)=\int p(z,x) dz$. This integral, called the \textit{evidence}, in some cases requires exponential time to be evaluated since it considers all configurations of latent variables. In other instances, it is unavailable in closed form \cite{vi_review}.

Variational Inference (VI) copes with this kind of problem by minimizing the Kullback-Leibler (KL) divergence between the true posterior distribution $p(z|x)$ and a parametric function $q(z)$, which is chosen among a set of densities $\Im$ \cite{vi_review}. This set of densities is parameterized by \textit{variational parameters} and they should be flexible enough to capture a density close to $p(z|x)$ and simple for efficient optimization. The parametric density which minimizes the KL divergence is 
\begin{equation}
q^*(z)=\argmin_{q(z)\in\Im} KL[q(z)||p(z|x)].
\label{KL_argmin}
\end{equation}
Unfortunately, Equation (\ref{KL_argmin}) cannot be optimized directly since it requires computing a function of $p(x)$. To see this, let us expand the KL divergence term using the Bayes' theorem and noting that $p(x)$ does not depend on $z$
\begin{align}
KL[q(z)||p(z|x)] =& E_{z \sim q}[log \ q(z) - log \ p(z|x)] \nonumber \\
=&  E_{z \sim q}[log \ q(z) - log \ p(x,z)]+log \ p(x). 
\label{KL_intrac}
\end{align}
Given that Equation (\ref{KL_intrac}) cannot be optimized directly, VI optimizes the alternative objective function 
\begin{align}
E_{z \sim q}[log \ p(x,z) - log \ q(z)]= & E_{z \sim q}[log \ p(z) + log \ p(x|z) - log \ q(z)] \nonumber \\
=& E_{z \sim q}[log \ p(x|z)]-KL[q(z)||p(z)] \nonumber \\
=& ELBO.
\label{ELBO}
\end{align}
From Equations (\ref{KL_intrac}) and (\ref{ELBO}) we have that
\begin{align}
log \ p(x)  = KL[q(z)||p(z|x)] + ELBO.
\end{align}
Since the KL divergence is non-negative, the expression in Equation (\ref{ELBO}) is called \textit{the evidence lower bound} (ELBO). Noting that the ELBO is the negative KL divergence in Equation (\ref{KL_intrac}) plus the constant term $log \ p(x)$, it follows that maximizing the ELBO is equivalent to minimizing Equation (\ref{KL_argmin}). 

The ELBO gives important information about the optimal variational density. The term $KL[q(z)||p(z)]$ encourages variational densities to be close to the prior distribution, while the term $E_{z \sim q}[log \ p(x|z)]$ encourages densities that place their mass on configurations of the latent variables that explain the observed data. The interested reader is referred to \cite{vi_review,doersch2016tutorial} for further details. 

\subsection{Variational Parameters Estimation}
\begin{figure}[t!]
    \centering
	\includegraphics[scale=0.52]{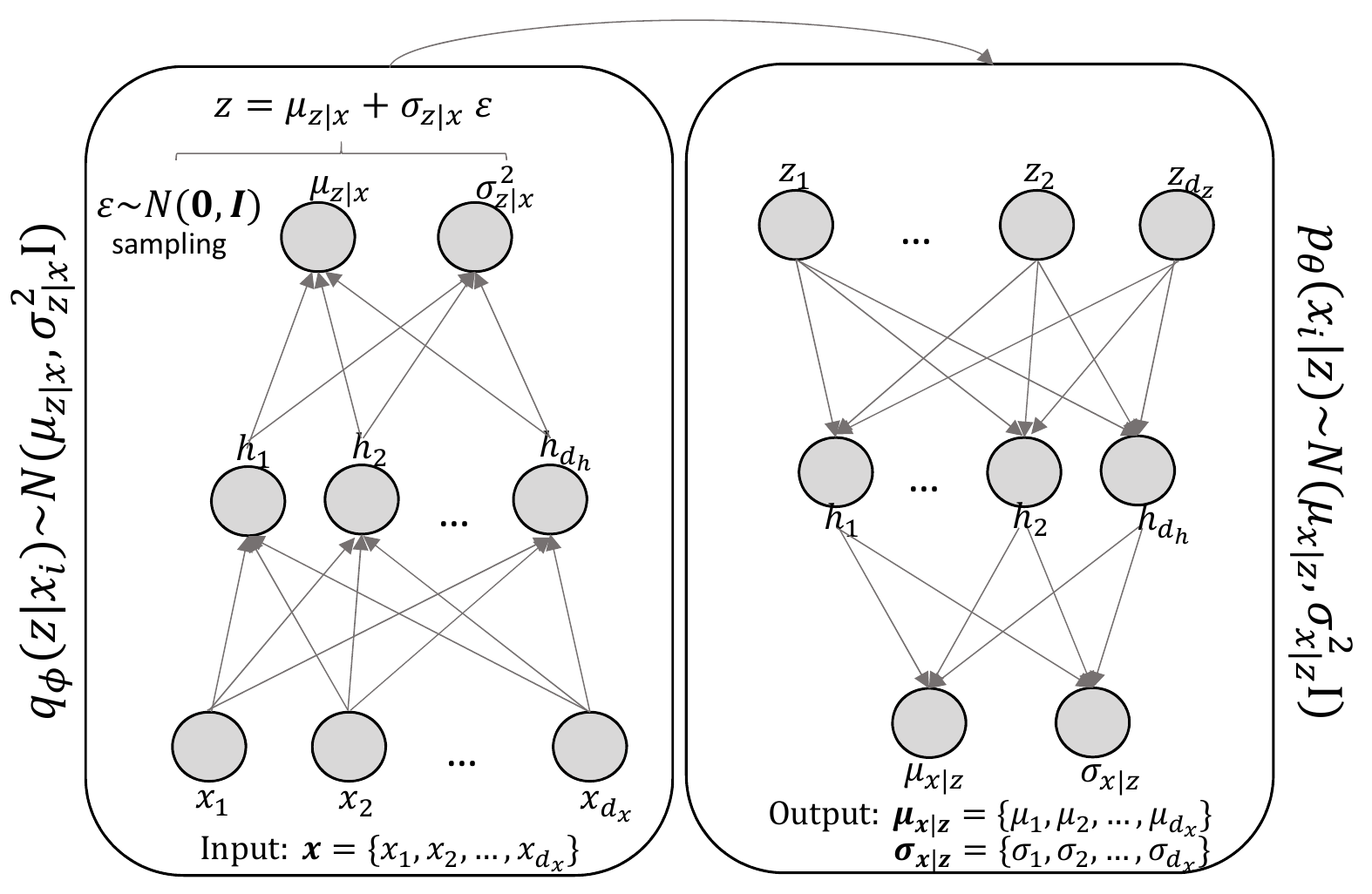}	
	\caption{Graphical representation of the AEVB algorithm. The feedforward neural network to the left corresponds to the probabilistic encoder $q_{\phi}(z|x_i)$ where $x_i \in \mathbb{R}^{d_x}$ is the network input. The output of the network are the parameters in \smash{$q_{\phi}(z|x_i) \sim \mathcal{N}(\bm{\mu}_{z|x_i}^{}, \mathbf{\sigma}_{z|x_i}^2 \mathbf{I})$}. Note that $\epsilon \sim \mathcal{N}(0,\mathbf{I})$ is drawn outside the network in order to use gradient descent and backpropagation optimization techniques. Similarly, the feedforward network to the right corresponds to the probabilistic decoder $p_{\theta}(x_i|z)$. In this case, the input are the latent variables $z \in \mathbb{R}^{d_z}$ and the network output are the parameters in \smash{$p_{\theta}(x_i|z) \sim \mathcal{N}(\bm{\mu}_{x_i|z}^{}, \mathbf{\sigma}_{x_i|z}^2 \mathbf{I})$}. The reconstruction is given by $\tilde{x}=\bm{\mu}_{x_i|z}$. For readability purposes we do not specify the parameters $\phi,\theta$ in the networks. However, these parameters are represented by the lines joining the nodes in the networks plus a bias term attached to each node. }
	\label{vae_diagram}
\end{figure}

The Variational Autoencoder (VAE), just as other autoencoders, uses feedforward neural networks to reproduce the input data. However, in this case, the encoder and decoder are probabilistic functions, see Figure (\ref{vae_diagram}). Specifically, the VAE approximates the true intractable posterior $p_{\theta}(z|x)$ in Equation (\ref{true_post}) by using the parametric approximation $q_{\phi}(z|x)$, implying two important distinctions in the way VAE solves the optimization problem presented in Section (\ref{sec_vi}). 

First, the parametric approximation is conditioned on the data $x$. The reason is that we are only interested in configurations of $z$ which are likely to reconstruct the data. Hence, by explicitly conditioning on $x$ we hope that the $z$ space is smaller, compared to the prior $p(z)$, and that it contributes to the term $E_{z \sim q}[log \ p(x|z)]$ in the ELBO \cite{doersch2016tutorial}. However, \cite{kingma2013auto} emphasizes that $q_{\phi}(z)$ can also be used as the parametric approximation for $p_{\theta}(z|x)$.

Second, we introduce the parameters $\theta,\phi$ in the densities $p$ and $q$ respectively. The reason is that the VAE estimates the expectation and covariance matrix of these density functions by training multilayer perceptron (MLP) networks. These networks learn these two parameters by backpropagating the ELBO with respect to all weights and bias in the MLP networks which are denoted by $\theta,\phi$.

Hence, assuming we have a set of i.i.d vectors $\{x_i,...,x_n\}$, $\theta$ and $\phi$ are learned by optimizing 
\begin{equation}
    ELBO(\theta,\phi,x_i) =  E_{z \sim q}[log \ p_{\theta}(x_i|z)]-KL[q_{\phi}(z|x_i)||p_{\theta}(z)]. 
    \label{alt_elbo}
\end{equation}
Note that $q_{\phi}(z|x_i)$ generates codes given $x_i$ and $p_{\theta}(x_i|z)$ converts those codes into its original representation. Hence, the former is also referred as probabilistic encoder and the latter as probabilistic decoder.

The Auto Encoding Variational Bayesian (AEVB) algorithm presented in \cite{kingma2013auto} estimates the parameters $\theta,\phi$ jointly using multilayer perceptron (MLP) networks. Therefore, assuming the prior is $p_{\theta}(z) \sim \mathcal{N}(\mathbf{0},\mathbf{I})$ and both the likelihood function $p_{\theta}(x_i|z)$ and the approximated posterior $q_{\phi}(z|x_i)$ are multivariate Gaussian with diagonal covariance $\mathcal{N}(\bm{\mu},\sigma^2\mathbf{I})$, the decoder and encoder share the same parameters as follows
\begin{align*}
h =& tanh(W_1 x_i + b_1), \\
\mu =& W_2 h + b_2, \\
log \ \sigma^2 =& W_3 h + b_3, \\
z_i =& \mu + \sigma \epsilon, 
\end{align*}
where $\epsilon \sim \mathcal{N(\mathbf{0},\mathbf{I})}$. Hence, $\phi =\{W_1,W_2,W_3,b_1,b_2,b_3\}$ and $\theta =\{W_4,W_5,W_6,b_4,b_5,b_6\}$ and when the MLP is used as decoder, $x_i$ is replaced by $z_i$ as the input data for the network. It is worth mentioning that the latent variable $z$ has been reparametrized as a deterministic and differentiable system. The reason is that we need to backpropagate the term $E_{z \sim q}[log \ p_{\theta}(x_i|z)]$ in Equation (\ref{alt_elbo}). Without the reparametrization, $z$ would be inside a sampling operation which cannot be propagated. This means that the AEVB algorithm actually takes the gradient of $E_{\epsilon \sim \mathcal{N(\mathbf{0},\mathbf{I})}}[log \ p_{\theta}(x_i|z_i= \mu_i + \sigma_i \epsilon )]$. The proof of this result can be found in \cite{kingma2013auto}. 


\section{Segmentation in the Latent Space}\label{sec_segmentation}
The idea behind the segment-based credit scoring is to identify segments with different propensity to fall into financial distress. The presumption is as follows. The endogenous variables describing the likelihood of whether a borrower would pay back a loan can be different from group to group. Besides, the degree of influence of a given variable can also vary from group to group. Further, given these between-groups differences, building segment-based classification models should increase the correct identification of customers who can fulfill their financial obligations and those who fail. 

To identify customer segments with different propensity to fall into financial distress, we take advantage of the term $E_{z \sim q}[log \ p(x|z)]$ in the evidence lower bound (ELBO) of the  Auto Encoding Variational Bayesian (AEVB) algorithm. As already mentioned in section (\ref{sec_vi}), $E_{z \sim q}[log \ p(x|z)]$ generates codes in the latent space which are likely to generate the input data. Hence, by transforming the input data, we can indirectly steer configurations in the latent space. Therefore, we do not use the VAE as a generative model or as a classification model, but as a novel clustering approach using the codes generated in the latent space.

To quantify customers propensity to fall into financial distress, let us first define the ground truth class
\begin{equation}
y=
    \left\{
    \begin{array}{ll}
      1 & if \ 90+dpd \\
      0 & otherwise,
    \end{array} 
    \right. \
    \label{default_flag}
\end{equation}
where \textit{90+dpd} stands for \textit{at least 90 days past due} and refers to customers' payment status. In the case of credit scoring, banks know the ground truth class of existing customers after the performance period is over \footnote{The performance period is the time interval in which if customers are at any moment 90+dpd, then their ground truth class is $y=1$. Frequently, 12 and 24 months are time intervals used for the performance period. Further, the performance period starts at the moment an applicant signs the loan contract.}. This variable can be used to build predictive models for new cases where the class label is unknown. There is some flexibility in the definition of the ground truth, and hence it may vary from bank to bank. However, it is common to use the above definition because it is aligned with the default definition in the Basel II regulatory framework \cite{anderson2007credit}. 

Further, let $C_j = \{c_1,c_2,...,c_n\}$ be the \textit{j'th} set of customers whom the ground truth class  $Y_j = \{y_1,y_2,...,y_n\}$  is known. Hence, 
\begin{equation}
    dr_{C_j} = \frac{\sum_i^n c_i[y_i=1]}{n},
    \label{eq_diffdr}
\end{equation}
where $[\cdot]$ is the Iverson bracket, can be referred to as the default rate of $C_j$ given that the ground truth class is aligned with Basel II.

Given that $dr_{C_j}>dr_{C_l}$, we say that $C_j$ has higher risk profile compared to ${C_l}$. In other words, customers in $C_j$ have, on average, higher probability of default $Pr(y=1)$, and the way we estimate $Pr(y=1)$ is using customers data. Therefore, in order to identify $k$ segments with a different propensity to fall into financial distress, we need to find segments where the average probability of default is different from the rest of groups. Mathematically,
\begin{equation}
    \frac{\sum_i^{n_{C_j}} Pr(y_i=1|x_i,C_j)}{n_{C_j}} \neq \frac{\sum_i^{n_{C_l}} Pr(y_i=1|x_i,C_l)}{n_{C_l}}, \quad for \quad j,l = 1,2,...,k, and \ j\neq l.
    \label{eq_diffdp}
\end{equation}
Now it is clear that the data transformation $f(x)$ that we are looking for, should incorporate the ground truth class $y$, i.e.  $f(x|y)$, given that we are interested in finding groups with a different average probability of default $Pr(y=1|x)$. In this way, the latent space in the VAE generates codes that also contain information about $y$. Otherwise, the codes will fail to reproduce $f(x|y)$.  

It turns out that there exists a function $f(x|y)$ in the logistic regression and the Naive Bayes literature which explains $Pr(y=1|x)$ as follows
\begin{align}
    \frac{Pr(y=1|x)}{1-Pr(y=1|x)}&=\frac{\frac{Pr(y=1)Pr(x|y=1)}{Pr(x)}}{\frac{Pr(y=0)Pr(x|y=0)}{Pr(x)}} \nonumber \\
    &=\frac{Pr(y=1)Pr(x|y=1)}{Pr(y=0)Pr(x|y=0)} \nonumber \\
    log \frac{Pr(y=1|x)}{1-Pr(y=1|x)}&=log \frac{Pr(y=1)}{Pr(y=0)}+log \frac{Pr(x|y=1)}{Pr(x|y=0)}.
    \label{woe_eq}
\end{align}
The second term in the right-hand-side of Equation (\ref{woe_eq}) is denoted Weight of Evidence (WoE) \cite{anderson2007credit,siddiqi2012credit}. In this setup, the prior log odds $\frac{Pr(y=1)}{Pr(y=0)}$ are constant, so the left hand side increases as the WoE increases. Originally, WoE was introduced by Irving John Good in 1950 in his book \textit{Probability and the Weighing of Evidence}.

The way to estimate $log \frac{Pr(x_j|y=1)}{Pr(x_j|y=0)}$, given that $x_j$ is continuous and implementing a fine classing of the WoE (see chapter 16.2 in \cite{anderson2007credit} or exhibit 6.2 in \cite{siddiqi2012credit}), is by creating $B_1, B_2, ..., B_k$ bins on the \textit{j'th} feature $x_j$. In the case of categorical variables, the different categories are already these bins. Hence, the WoE for the \textit{k'th} bin is  
\begin{equation}
    WoE_{kj} = log \frac{Pr(x_j \in B_k|y=1)}{Pr(x_j \in B_k|y=0)} = log \frac{\big(\sum_j x_j \in B_k [y=1]\big)/n_{x_j}}{\big(\sum_j x_j \in B_k [y=0]\big)/n_{x_j}},
\end{equation}
where $[\cdot]$ is the Iverson bracket. Note that the number of bins $B_k$ can be different for different features. 

It is worth mentioning that bank analysts implemented a coarse classing of the WoE that we use to train the VAE. The coarse classing approach merges bins, obtained in the fine classing step, with similar WoE (see chapter 16.2 in \cite{anderson2007credit} or exhibit 6.2 in \cite{siddiqi2012credit}). For continuous variables, only adjacent bins are merged, while for categorical variables non-adjacent bins can also be combined.

We use the WoE transformation of the input data $x$ to tilt the latent space in VAE towards configurations which encapsulate the propensity to fall into financial distress. Remember that the goal is to reveal groups that satisfy Equation (\ref{eq_diffdp}). Further, Equation (\ref{woe_eq}) shows that the WoE explains the posterior log odds which are just a function of the default probability in Equation (\ref{eq_diffdp}). Hence, the presumption is that given that the AEVB algorithm has converged to the optimal variational density $q^*(z)$, the latent space should encapsulate the probability of default. Otherwise, the reconstruction would have failed, and the algorithm would not have converged to $q^*(z)$ in the first place. 

\begin{figure}[t!]
    \centering
	\includegraphics[scale=0.45]{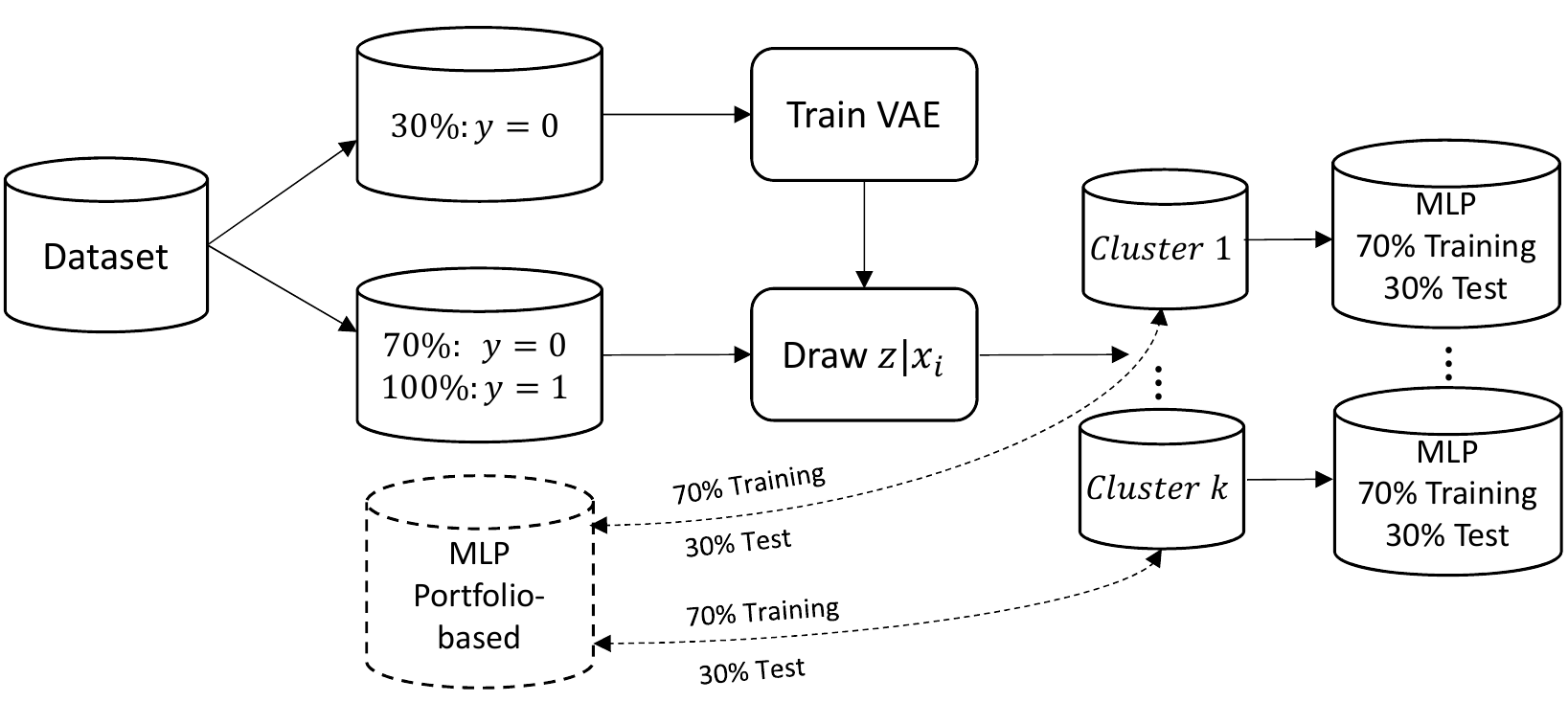}	
	\caption{Graphical representation of the development methodology where we use 30\% of the majority class ($y=0$) data for training the VAE. Once the VAE is trained, it is used to generate the latent variables $z$ for the remaining data, i.e., 70\% of the majority and 100\% of the minority class ($y=1$). Based on the clusters in the latent space, we train MLP classifier models using a classical 70\%-30\% partition for training and testing the model respectively. The model performance of these segment-based MLP models is compared against a portfolio-based MLP model where no segmentation is considered (the dashed box denotes this model).}
	\label{data_partition}
\end{figure}

\section{Experiments and Results}\label{sec_results}
\subsection{Data description}\label{data_sum}
We use three datasets in our experiments; a Norwegian and a Finnish car loan dataset from Santander Consumer Bank Nordics and the public dataset used in the Kaggle competition \textit{Give me some credit}\footnote{Website \url{https://www.kaggle.com/c/GiveMeSomeCredit}}. These datasets show applicants' status, financial and demographic factors, at the time of application. Further, the ground truth class is known after the applicant became a customer and at the end of the performance period. The performance period for the real datasets is 12 months, while for the public dataset it is 24 months. More details about the datasets can be found in Tables (\ref{tbl_summary_dta}), (\ref{tbl_variables_no}) and (\ref{tbl_variables}).

The following sections show different experiments to assess if i) we are able to steer configurations in the latent space of the VAE through a meaningful data transformation, and ii) if the VAE clustering structure is well suited for segment-based credit scoring and may perform better than the portfolio-based approach. The focus of the analysis is on the Norwegian car loan dataset since we have relatively more insight into this dataset.

\subsection{Clustering in the latent space}\label{sec_clustering_inZ}
It is worth mentioning that the VAE could be trained using both the minority and the majority class. However, given the high class imbalance in the datasets under analysis, incorporating data from the minority class into the VAE training means that we need to train and test the classifier model with fewer customers from this class. This is, in general, not the desired scenario. Similarly, the number of customers from the minority class that could have been added to train the VAE would not have changed our results considerably. In cases where there is enough data from both classes, it is definitively recommended to use both classes to reveal clusters in the latent space and see if there are benefits by doing so. 

Hence, we use 30\% of the majority class $y=0$ to train the VAE. During the training phase, we generate the latent space for the remaining data, i.e. 70\% of the majority class data and 100\% of the minority class data $y=1$, see Figure (\ref{data_partition}). Based on the cluster quality of the generated latent space together with the optimization of the ELBO, we select the optimal network architecture as well as the stopping criteria.

\begin{algorithm}[t!]
\SetAlgoLined
\SetKwInOut{Input}{Input}
\SetKwInOut{Output}{Output}
\Input{Z,$n_{min}$,$\alpha$,$\rho$}
\Output{cluster labels}
pending\_data = \{Z\} \;
 labels = ones(length(Z)) \;
 \While  {EOF(pending\_data)==FALSE} { 
      \For {item in pending\_data} {
      labels = HierarchicalAlgorithm(pending\_data[item], k = 2) \;
      get centroids c1 and c2 \;
      split pending\_data[item] into C1 and C2 using labels  \;
      \If {$n1 > n_{min}$ AND $n2 > n_{min}$ AND  $||c1 - c2|| > \rho$}
      {
      update labels \;
      pending\_data.append = \{C1,C2\}
      }
      }
  }
  return labels
 \caption{Labeling the VAE clustering structure in the latent space.}
 \label{alg_1}
\end{algorithm}

This means that the focus during training is both the optimization of the ELBO and the clustering structure in the latent space. In other words, the selected architecture for the VAE is the one that has converged to the optimal variational density $q^*(z)$, and that generates clusters in the latent space with low within cluster scatter and which are well spread. This cluster property is measured using the Calinski-Harabaz (CH) index introduced in \cite{calinski1974dendrite}. Furthermore, the clusters should have considerably different default rates. This cluster characteristic is measured using the average maximum between-cluster default rate (BCDR)
\begin{align}
    dr_{C_i}^* &= \max_{j \neq i} ||dr_{C_i} - dr_{C_j}||, \nonumber \\
    BCDR &= \frac{1}{k}\sum_i^k dr_{C_i}^*,
\end{align}
where $dr_{C_i}$ is the default rate of the \textit{i'th} cluster, $||\cdot||$ is the Euclidean distance, and $k$ is the total number of clusters in the latent space. Therefore, the quality of the clusters in the latent space is high when both the CH index and the BCDR index are high.

We assign cluster labels in the latent space using the following approach. At each \textit{j'th} epoch, we estimate the expectation of the latent variable $z$ 
\begin{equation}
    E_q[z|x_i]=\int_{-\infty}^{\infty}z q_{\phi}(z|x_i)dz\approx\frac{1}{n}\sum_{j=1}^n z_j,
\end{equation}
given the input vector $x_i$. We use the the expectation of $n$ latent variables to average out uncertainty and obtained relatively more compact clusters since $q_{\phi}(z|x_i) \sim \mathcal{N}(\mathbf{\mu_{z|x_i}}, \mathbf{\sigma_{z|x_i}^2}\mathbf{I})$. Further, using the expectation also has the advantage that customers are assigned to the same cluster each time we generate their latent variables. In our experiments, we found robust results for $n=100$.

\subsubsection{Norwegian car loan dataset}
Now that the latent space is generated, we use hierarchical clustering iteratively to assign cluster labels, always partitioning the data into two clusters at each iteration, see Algorithm (\ref{alg_1}). The parameters $n_{min}$ (minimum cluster size) and $\rho$ (minimum Euclidean distance between clusters) in the algorithm are selected based on data. The only requirement for these parameters is that the algorithm must preserve the clustering structure in the latent space of the VAE. In other words, this approach is only an automated mechanism to label the clustering structure in the latent space. In this way, we can evaluate the clustering structure when we optimize the ELBO and we can select the optimal architecture based on quantitative grounds.

Finally, we use grid-search to select the hyper-parameters of the VAE. We tested different architectures varying the number of hidden units, the number of hidden layers, the dimensionality of the latent space and the learning rate. For each of these architectures, we evaluate the quality of the clustering structure and the optimization of the ELBO. Table (\ref{all_archs}) shows some of the architectures we tested. It is important to mention that other complex architectures, e.g., two or more hidden layers and higher dimensions of the latent space, were also tested. However, for the datasets we are analyzing, the AEVB algorithm takes longer to converge to the optimal variational density $q^*(z)$ for high dimensional spaces in the latent space. On the other hand, for the Kaggle dataset adding more hidden layers makes the algorithm converge faster, but the resulting clustering structure in the latent space contains only two clusters. The algorithm also converges faster for the Finnish dataset if we use more than one hidden layer, yet the resulting clustering structure contains four clusters which is not ideal for this dataset as we will see later on. See Figure (\ref{elbo_op}) for more details.

\begin{figure}[t!]
    \centering
	\includegraphics[scale=0.35]{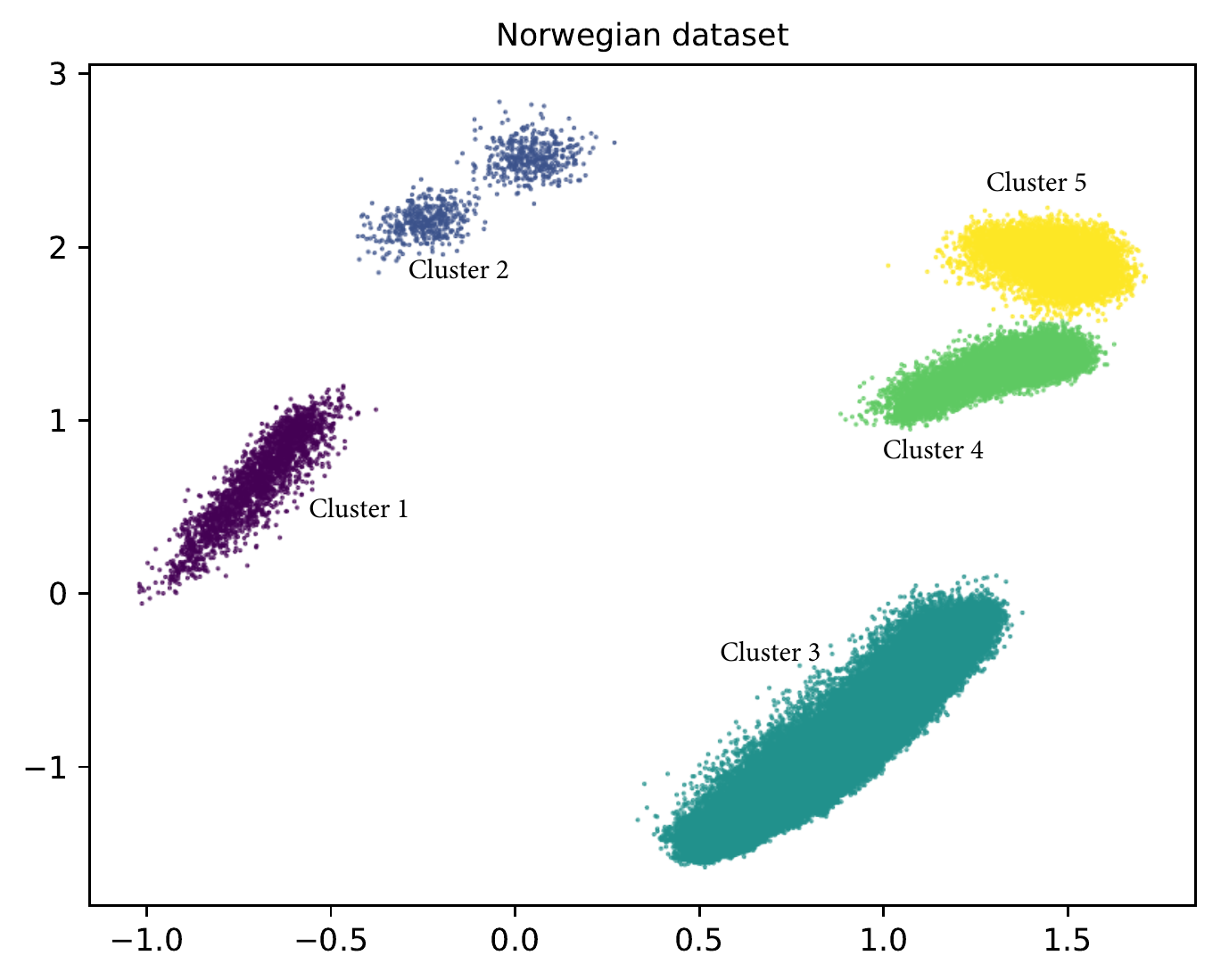}	
	\caption{Cluster structures in the VAE latent space for the Norwegian car loan dataset.}
	\label{k_in_z}
\end{figure}

\begin{table}[t!]
\centering 
\begin{adjustbox}{width=8cm,totalheight=2cm}
\begin{tabular}{ |c|c|c|c|c|}
\hline
\multicolumn{5}{|c|}{\textbf{\large{Norwegian car loan}}} \\
\hline
Cluster & Default rate & Avg. Probability of Default & Nr. Customers & Nr. $y=1$ \\
\hline
1 & 5.30\%   & 3.70\% & 2 206   & 117  \\
2 & 11.24\%  & 9.99\% & 774     & 87   \\
3 & 1.39 \%  & 0.94\% & 109 969 & 1539 \\
4 & 2.89\%   & 2.27\% & 11 450  & 331  \\
5 & 5.30\%   & 4.32\% & 9 106   & 483  \\
\hline
\end{tabular}
\end{adjustbox}
\caption{Risk profiles, default rates $dr_{C_j}$ and average probability of default $Pr(y_i=1|x_i,C_j)$, for the different clusters $C_j$ in the VAE two-dimensional latent space.}
\label{tbl_k_results}
\end{table}

The clustering structure found for the Norwegian dataset is shown in Figure (\ref{k_in_z}), while Table (\ref{tbl_k_results}) shows the associated risk profiles. The first important result to highlight is that the WoE transformation reveals well-defined clusters in the latent space. Further, with the exception of cluster 1 and 5, default rates (dr) and the average probability of default $Pr(y=1|x, C_j)$ for most clusters are considerably different. Note that $Pr(y=1|x, C_j)$ is estimated using the multilayer perceptron models that we will present in Section (\ref{sec_segbasedmlp}). Given the few customers from the minority class, we decided not to set aside a dataset for probability calibration. Therefore, the average probabilities of default are not well calibrated to the default rates.  

In general, the Norwegian car portfolio has a low-risk profile (the dataset contains only 2 557 customers from the minority class). Hence, it makes sense that about 82\% of all customers are in cluster 3, the cluster with the smallest default rate and average probability of default, while about 18\% of the customers are in clusters with relatively high-risk profiles. 

\begin{figure}[t]
    \centering
	\includegraphics[scale=0.5]{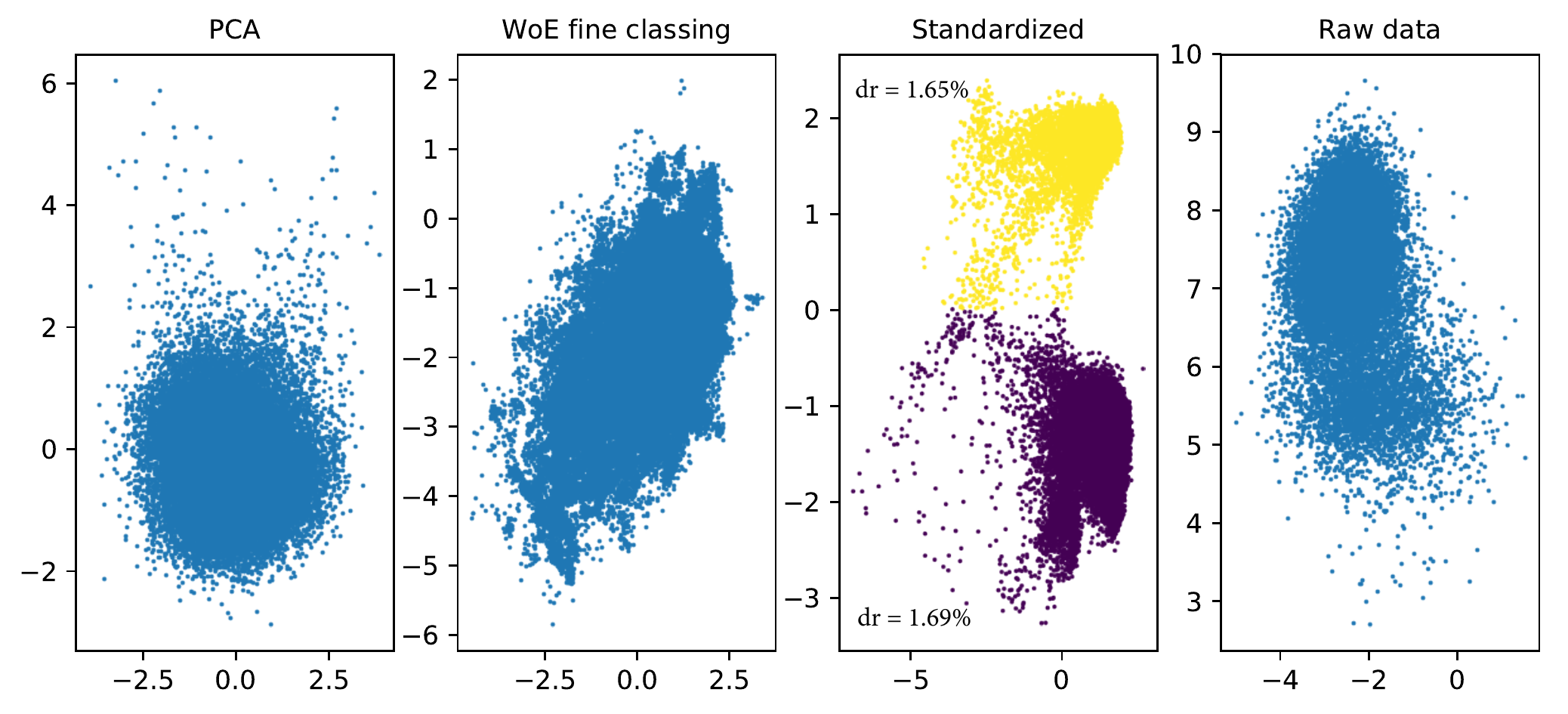}	
	\caption{Latent space for four different data transformation for the Norwegian dataset. The left panel shows a PCA transformation (preserving the original data dimensionality). The second panel shows the latent space for the fine classing WoE transformation. The third panel shows the latent space for the standardized data, and finally, the right panel shows the latent space for the raw data. Standardizing the data reveals two clusters in the latent space. However, these clusters have practically the same default rate ($dr$). The other three transformations do not show any clustering structure.}
	\label{four_trans}
\end{figure}

Now we want to show that the (coarse classing) WoE has valuable information for revealing clusters in the latent space of the VAE. Hence, we use different data transformations, and for each of these transformations we train a VAE with a two-dimensional latent space $z \in \mathbb{R}^2$, i.e. for each data transformation, we reduce the original number of input features into two dimensions in the latent space of the VAE. Specifically, we generate the latent space for the following data transformations:
\begin{enumerate}
    \item Scaling with PCA: The input data is transformed using principal component analysis with all the principal components, i.e. there is no dimensionality reduction.
    
    \item Scaling with Standardization: The input data is standardized by removing the mean and scaling to unit variance.
    
    \item Scaling with fine classing WoE: The input data is transformed into WoE by creating bins with an approximately equal number of customers, i.e. no coarse classing is done by bank analysts.    
    
    \item Input data: Raw data without scaling.
\end{enumerate}
Figure (\ref{four_trans}) shows the resulting latent spaces for the data transformations explained above. Interestingly, three of these transformations do not show any clustering structure at all, and for the last, the clusters have practically the same default rate. Hence, by identifying appealing data transformations, such as the WoE with coarse-tuning, and leveraging the properties of the ELBO in the VAE, it is possible to steer configurations in the latent space. In our particular case, these configurations are well-defined clusters with considerably different risk profiles. 

\subsubsection*{Cluster interpretation}
For the bank industry, it is essential to understand which features that are most important for the clustering result. To investigate this issue, we adopt the salient dimension methodology presented in \cite{azcarraga2005extracting}. This approach identifies features whose values are statistically significant in different clusters. These are called salient dimensions. 

Let $v$ be the \textit{v'th} dimension of the \textit{i'th} vector $x_{i,v}$, where $x \in R^{\ell}$. Further let $\Phi_{in}(k)$ be the set of in-patterns (within cluster $k$) and $\Phi_{out}(k)$ be the set of out-patterns (not within cluster $k$). Then compute the mean input values  
\begin{align}
    \mu_{in}(k,v) &= \frac{\sum_{x_i \in \Phi_{in}(k)} x_{i,v} }{|\Phi_{in}(k)|}, \\
    \mu_{out}(k,v) &= \frac{\sum_{x_i \in \Phi_{out}(k)} x_{i,v} }{|\Phi_{out}(k)|},
\end{align}
where $|\{\cdot\}|$ returns the cardinality of $\{\cdot\}$. Further, compute the difference factors 
\begin{equation}
    df(k,v)=\frac{\mu_{in}(k,v)-\mu_{out}(k,v)}{\mu_{out}(k,v)},
\end{equation}
and their mean and standard deviations
\begin{align}
    \mu_{df}(k) &= \frac{1}{\ell}\sum_v^\ell df(k,v),  \\
    \sigma_{df}(k) &= \sqrt{\sum_v^\ell\big(df(k,v)-\mu_{df}(k)\big)^2/\ell}.
\end{align}
Finally, we say that the \textit{v'th} feature in cluster $k$ is a salient dimension if
\begin{equation}
    df(k,v) \leq \mu_{df}(k) - s.d. \ \sigma_{df}(k),
\end{equation}
or
\begin{equation}
    df(k,v) \geq \mu_{df}(k) + s.d. \ \sigma_{df}(k), 
\end{equation}
where $s.d.$ is the number of standard deviations to be used. The value for $s.d$ is defined based on the dataset. We use $s.d.=1$ for all three datasets under analysis.

Now let us explain the clustering structure in Figure(\ref{k_in_z}) together with the risk profiles in Table (\ref{tbl_k_results}) using the salient dimensions for the Norwegian car loan dataset in Table (\ref{tbl_saldim}). 

The first interesting result is the pattern of the latent variables for clusters 1 and 5 (both clusters have a default rate equal to $5.30\%$) which are located on opposite sides of the two-dimensional space. The salient dimension \textit{MaxBucket12} in cluster 1 shows that about $70\%$ of the customers have between 30 and 60 dpd at the moment they applied for the loan, i.e. they are existing customers applying for a new loan. Actually, all customers in cluster 1 are existing customers who are at least 30 days past due. On the other hand, about $51\%$ of the customers in cluster 5 are new customers. Cluster 2 is also composed of existing customers only. Hence, it seems like clusters without existing customers are placed on the left side of Figure(\ref{k_in_z}).

Now let us see what characterizes cluster 3 which is the cluster with the lowest default rate. Looking at the salient dimension \textit{DownPayment\%}, we can see that the average value in this cluster is about $20\%$, while for the rest of the clusters the average down payment is less than $12\%$. When it comes to the salient dimension \textit{AgeObject}, we can observe that about $35\%$ of the customers in this cluster have relatively new cars, while for the rest of clusters the average percentage of customers with new cars is about $23\%$.

Finally, cluster 2 has the highest default rate, which can be easily explained by its salient dimension \textit{DownPayment\%}. About $93\%$ of customers in this cluster are between 1 and 90 days past due. While the percentage of customers in the other clusters with the same number of days past due is only $15\%$. 

There are five factors where segmentation can play a significant role in the bank industry (see Section (\ref{sec_literature})). Specifically, given the above discussion and that the results are aligned with business expectations, the clusters identified for the Norwegian car loan dataset can be useful for marketing, customer and model fit factors. 

\begin{figure}[t!]
    \centering
	\includegraphics[width=8cm,height=4cm]{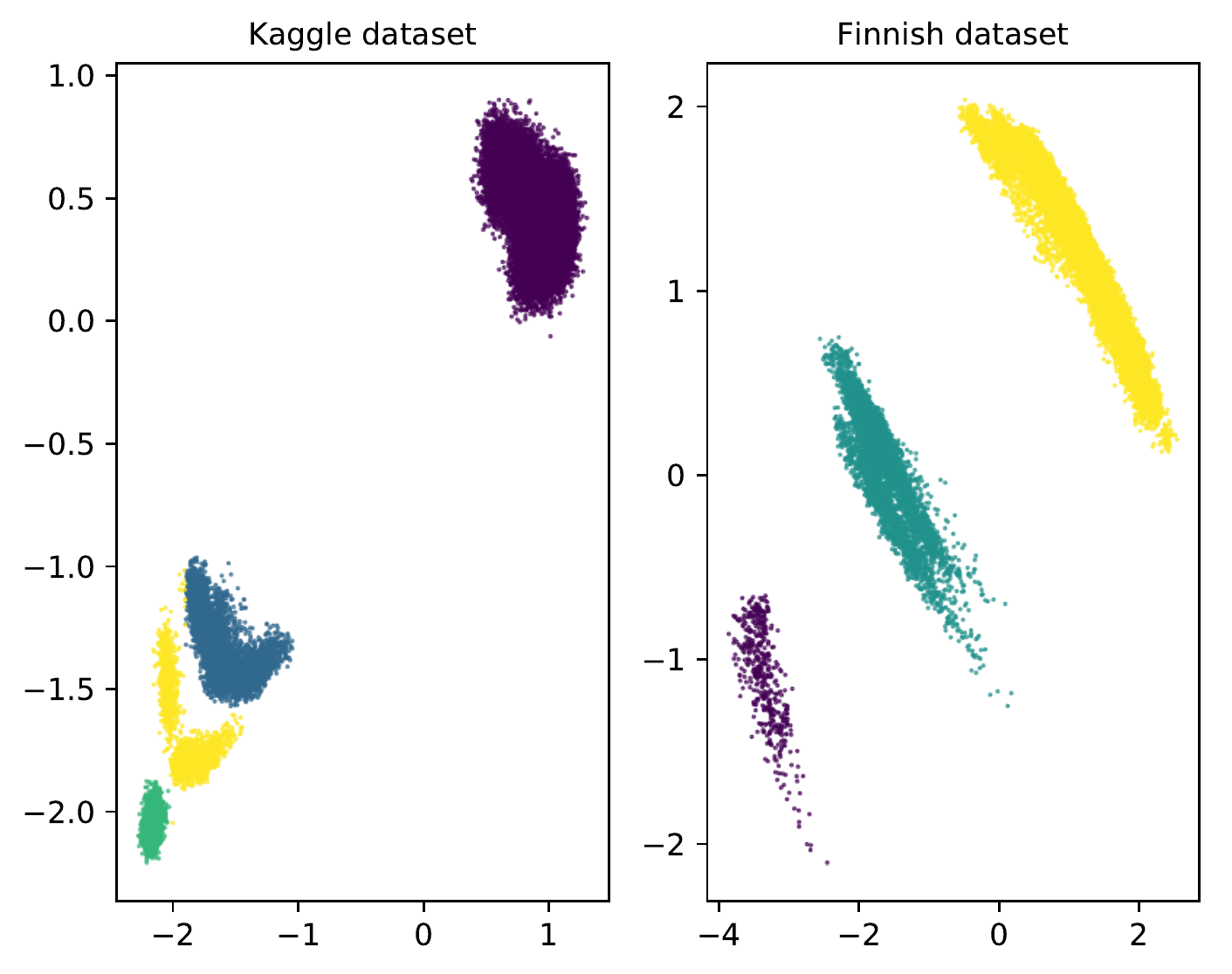}	
	\caption{Clustering structures for the Kaggle and Finnish car loan dataset in the latent space of the VAE.}
	\label{k_in_z_k_fi}
\end{figure}

\begin{table}[t!]
\centering 
\begin{adjustbox}{width=13cm,totalheight=1.71cm}
\begin{tabular}{ |c|c|c|c|c|c|c|c|c|}
\hline
 & \multicolumn{4}{c|}{\textbf{\large{Kaggle}}} & \multicolumn{4}{c|}{\textbf{\large{Finnish car loan}}} \\
\cline{1-9}
Cluster & Default rate & Avg Probability of Default & Nr. Customers & Nr. $y=1$ & Default rate  & Avg Probability of Default  & Nr. Customers & Nr. $y=1$\\
\hline
1 &  5.47\%   & 5.54\%  & 97 434  &  5 327   & 5.93\%   & 4.80\% & 438    & 26     \\
2 &  33.13\%  & 32.62\% & 6 121   &  2 028   & 3.76\%   & 3.29\% & 6 067  & 228    \\
3 &  55.68\%  & 55.70\% & 2 026   &  1 128   & 0.91\%   & 0.37\% & 75 536 & 685    \\
4 &  63.58\%  & 63.22\% & 2 427   &  1 543   &          &        &        &        \\
\hline
\end{tabular}
\end{adjustbox}
\caption{Risk profiles, default rates $dr_{C_j}$ and average probability of default $Pr(y_i=1|x_i,C_j)$, for the different clusters $C_j$ in the two-dimensional latent space of the VAE.}
\label{tbl_k_results_k_fi}
\end{table}

\subsubsection{Kaggle and Finnish car loan dataset}
We now repeat the clustering procedure for the Kaggle and Finnish car loan datasets. Figure (\ref{k_in_z_k_fi}) shows the clustering structures in the VAE latent space. As for the Norwegian dataset, the clustering structures for these datasets are well defined, with four clusters for the Kaggle dataset and three clusters for the Finnish dataset. 

The risk profiles for these two datasets are shown in Table (\ref{tbl_k_results_k_fi}). We observe the same pattern as in the Norwegian clusters, where the majority of the customers are in the cluster with the smallest default rate and the smallest average probability of default. However, the clustering structure for the Kaggle dataset has some interesting results. About 10\% of the customers are in three clusters with very high-risk profiles. Another interesting aspect is that the estimated average probability of default is well-calibrated compared to their default rates. The relatively large number of customers from the minority class in each cluster (more than 1 100 in all clusters) probably explains this result.

Finally, salient dimensions for the Kaggle and Finnish car loan datasets can be found in Table (\ref{tbl_saldim}).

\subsection{Segment-based credit scoring}\label{sec_segbasedmlp}
Next we turn our attention to the classification task based on the clustering structures found in the previous section. The segment-based credit scoring approach is based on the presumption that the endogenous variables describing the likelihood of whether a customer would fall into financial distress can be different from group to group, given that these groups have, on average, different probability of default. In addition, the degree of influence of a given variable can vary from cluster to cluster. 

We compare the classification performance of the segment-based approach (one multilayer perceptron (MLP) model for each cluster) against the classical classification approach (only one MLP model for the whole dataset) using the H-measure to assess if there are performance gains in the segment-based approach. The main reason for using the H-measure is that it is well suited for highly class imbalanced datasets. Furthermore, comparing multiple models with the H-measure, it is done objectively since the H-measure specifies a common misclassification cost for all models. For further details please refer to \cite{hand2009measuring,hand2014better}. For purposes of comparison, we provide other well-known performance metrics, e.g., the area under the receiver operating characteristic curve (AUC), the Gini coefficient, and the Kolmogorov-Smirnov (KS) test.

\begin{table}[t!]
\centering 
\begin{adjustbox}{width=8cm,totalheight=6cm}
\begin{tabular}{ |c|c|c|c|c| } 
\hline
\multicolumn{5}{|c|}{\textbf{\large{Norwegian car loan}}} \\
\hline
Performance metric & Cluster & Segment-based & Portfolio-based & \textit{p}-value \\
\hline
\multirow{5}{*}{\textbf{{H-measure}}}   & 1 & 0.2774 & 0.2310   &  0.0509  \\
                                        & 2 & 0.1665 & 0.1453   &  0.4299  \\
                                        & 3 & 0.2174 & 0.2076   &  0.0854  \\
                                        & 4 & 0.1760 & 0.1471   &  0.0186  \\
                                        & 5 & 0.1302 & 0.1193   &  0.0931  \\
\hline
\multirow{5}{*}{\textbf{{AUC}}}         & 1 & 0.7688 & 0.7430   &          \\
                                        & 2 & 0.6791 & 0.6701   &          \\
                                        & 3 & 0.7756 & 0.7706   &          \\
                                        & 4 & 0.7395 & 0.7188   &          \\
                                        & 5 & 0.7021 & 0.6923   &          \\
\hline
\multirow{5}{*}{\textbf{{Gini}}}        & 1 & 0.5377 &  0.4860  &          \\
                                        & 2 & 0.3582 &  0.3402  &          \\
                                        & 3 & 0.5511 &  0.5412  &          \\
                                        & 4 & 0.4790 &  0.4377  &          \\
                                        & 5 & 0.4043 &  0.3846  &          \\
\hline                                        
\multirow{5}{*}{\textbf{{KS}}}          & 1 & 0.4648 &  0.4098  &          \\
                                        & 2 & 0.3441 &  0.3280  &          \\
                                        & 3 & 0.4318 &  0.4199  &          \\
                                        & 4 & 0.3821 &  0.3489  &          \\
                                        & 5 & 0.3410 &  0.3299  &          \\
\hline
\end{tabular}
\end{adjustbox}
\caption{Average model performance, for the segment-based credit scoring approach and for the portfolio-based approach, based on a 10-cross-validation approach. Note that best models are selected based on the H-measure.}
\label{tbl_performance_no}
\end{table}

The development dataset is composed of 70\% of the majority class and 100\% of the minority class (none of the data used for training the VAE is used in the classification exercise). Further, this sample is divided into the classical 70\%-30\%  split for training and testing the model, see Figure (\ref{data_partition}), and we make sure that the test datasets keep the original class ratio between the majority and minority class. It is worth mentioning that given the high class-imbalance in the datasets, we generate synthetic variables to train the MLP models. Then, we correct the bias in the estimated probability of default adopting the methodology presented in \cite{dal2015calibrating}. 

Further, due to the class imbalance problem we use a cross-validation approach in which at each \textit{k'th} iteration we: i) divide the development sample into 70\%-30\% for training and testing, ii) generate synthetic variables to balance the training dataset, and iii) train the MLP model using the balanced dataset and test it on the original class imbalance ratio. It is worth mentioning that when we train the portfolio-based MLP model, we make sure that the training and test datasets include exactly $k^{-1}$ times the number of customer in each \textit{k'th} cluster. In theory, cluster labels are unknown in the portfolio-based approach, but in practice not doing the sampling in this way, could potentially benefit performance in one cluster by harming the performance of another. Finally, for both the segment-based and portfolio-based approach we use ten cross-validations and grid-search for hyper-parameter tuning.

\subsubsection{Norwegian car loan dataset}
The classification results for the Norwegian car loan dataset are shown in Table (\ref{tbl_performance_no}). The performance of the segment-based approach is better than that of the portfolio-based approach for all clusters.  Note that the difference in H-measure between the two approaches is smallest for cluster 3, which may suggest that the largest cluster drives model performance in the portfolio-based approach. For clusters with high-risk profiles, there are large difference in performance, suggesting that the underlying risk drivers in such clusters are different. Hence, there seems to be a clear advantage to build different classifier models for groups of customers with distinct risk profiles. 

To get a better overview of model performance as measured by the H-measure, we conduct an unpaired t-test. This statistical test checks whether the average difference in model performance between the two approaches is significantly different from zero. The \textit{p}-values of the t-test for the Norwegian car loan dataset are shown in the third column of Table (\ref{tbl_performance_no}). It is not surprising that the difference in H-measures for cluster 2 are not significant, given the small cluster size and few customers from the minority class (only 87). The difference in H-measures for clusters 3 and 5 are significant at the 10\% level, while those for clusters 1 and 4 are significant at the 5\% level. Note that there are some concerns with the unpaired t-test and the methodology we used to build the MLP classifiers models: i) overlapping in the cross-validation datasets, ii) sample size (only ten cross-validations) and iii) outliers skew the test statistics and decrease the test's power \cite{demvsar2006statistical,dietterich1998approximate}. Further, the synthetic variables generated to train the MLP models are an extra source of variability that could impact the \textit{p}-values.

The H-measure, as well as the rest of performance metrics in Table (\ref{tbl_performance_no}), consistently rank the model performance of the segment-based approach on top of the portfolio-based approach. Hence, the insight provided by the t-test should be taken as informative and not as conclusive. 

\begin{table}[t!]
\centering 
\begin{adjustbox}{width=12cm,totalheight=5.5cm}
\begin{tabular}{ |c|c|c|c|c|c|c|c| } 
\hline
& &  \multicolumn{3}{c|}{\textbf{\large{Kaggle}}} & \multicolumn{3}{c|}{\textbf{\large{Finnish car loan}}} \\
\cline{1-8}
Performance metric & Cluster & Segment-based & Portfolio-based & \textit{p}-value & Segment-based & Portfolio-based & \textit{p}-value \\
\hline
\multirow{5}{*}{\textbf{{H-measure}}}   & 1 & 0.2918    &  0.2842  & 0.0410 & 0.1949 & 0.2388  & 0.4141  \\
                                        & 2 & 0.1145    &  0.1067  & 0.2230 & 0.0946 & 0.1001  & 0.5870  \\
                                        & 3 & 0.0828    &  0.0691  & 0.0883 & 0.1838 & 0.1817  & 0.8305  \\
                                        & 4 & 0.0765    &  0.0660  & 0.0626 &        &         &         \\
\hline
\multirow{5}{*}{\textbf{{AUC}}}         & 1 & 0.8047    &  0.8018  &    & 0.5916 & 0.6247  &          \\
                                        & 2 & 0.6680    & 0.6629   &    & 0.6491 & 0.6596  &          \\
                                        & 3 & 0.6284    & 0.6126   &    & 0.7366 & 0.7370  &          \\
                                        & 4 & 0.6270    & 0.6189   &    &        &         &          \\
\hline
\multirow{5}{*}{\textbf{{Gini}}}        & 1 & 0.6094 & 0.6036  &        & 0.1832    & 0.2495  &        \\
                                        & 2 & 0.3360 & 0.3258  &        & 0.2983    & 0.3193  &        \\
                                        & 3 & 0.2569 & 0.2253  &        & 0.4733    & 0.4741  &        \\
                                        & 4 & 0.2540 & 0.2379  &        &           &         &         \\
\hline                                        
\multirow{5}{*}{\textbf{{KS}}}          & 1 & 0.4711 & 0.4633  &        & 0.3290    & 0.3504  &        \\
                                        & 2 & 0.2550 & 0.2480  &        & 0.2797    & 0.3009  &        \\
                                        & 3 & 0.2218 & 0.1817  &        & 0.3735    & 0.3688  &        \\
                                        & 4 & 0.2067 & 0.2034  &        &           &         &         \\
\hline
\end{tabular}
\end{adjustbox}
\caption{Average model performance, for the segment-based credit scoring approach and for the portfolio-based approach, based on a 10-cross-validation approach. Note that best models are selected based on the H-measure.}
\label{tbl_performance_k_fi}
\end{table}

\subsubsection{Kaggle and Finnish car loan dataset}
Table (\ref{tbl_performance_k_fi}) shows the performance of the segment-based and portfolio-based approach for the Kaggle and Finnish car loan dataset. For the Kaggle dataset, the performance of the segment-based approach is better for all clusters. We can see the same pattern as for the Norwegian car loan dataset. The smallest performance gain of the segment-based approach is for the largest cluster, and there are larger differences in performance for the high-risk clusters.

For the Finnish dataset, the segment-based approach does not increase model performance. Looking at Table (\ref{tbl_k_results_k_fi}), we can see that the total number of customers from the minority class is only 939 for this dataset. Further, these customers are spread across the three different clusters, with 25, 228 and 685 instances in cluster 1, 2 and 3 respectively. It is challenging to train classifier models when the number of customers from the minority class is this small. 

\section{Conclusion}\label{sec_conclusion}
In this paper, we show that it is possible to steer configurations in the latent space of the Variational Autoencoder (VAE) by transforming the input data. Specifically, the Weight of Evidence (WoE) transformation encapsulates the propensity to fall into financial distress and the latent space in the VAE preserves this characteristic in a well-defined clustering structure. This result shows the powerful information embedded in the latent space, which has been documented in different research domains \cite{bioinformatics18, hou2017deep, latif2017variational,rampasek2017dr,bowman2015generating}. Further, the resulting clusters are characterized by having considerably different risk profiles measured by the default rate.

The Variational Autoencoder has the advantage of capturing non-linear relationships which are projected in the latent space. In addition, the number of clusters is clearly suggested by the latent space in the VAE and, for a low-dimensional latent space, they can be visualized. Furthermore, the VAE can generate the latent configuration of new customers and assign them to one of the existing clusters. The clustering structure in the latent space of the VAE can be used for marketing, customer, and model fit purposes in the bank industry, given that the clusters have considerably different risk profiles and their salient dimensions are business intuitive.

Finally, we show that for portfolios with a large number of customers in each cluster, as well as a sufficient number of customers from the minority class, the clusters in the latent space of the VAE can be used for segment-based credit scoring. According with the H-measure, the segment-based credit scoring approach performs better than the traditional portfolio-based credit scoring where only one classifier is trained for the whole customer portfolio. The fact that the segment-based credit scoring performs better compared to the portfolio-based credit scoring, specially in clusters with high-risk profiles, suggests that the underlying risk drivers in each cluster are different. 

\section*{Acknowledgements}
The authors would like to thank Santander Consumer Bank for financial support and the real datasets used in this research. This work was also supported by the Research Council of Norway [grant number 260205] and SkatteFUNN [grant number 276428].

\clearpage
\section*{Appendices}
\renewcommand{\thefigure}{A\arabic{figure}}
\setcounter{figure}{0}
\renewcommand{\thetable}{A\arabic{table}}
\setcounter{table}{0}

\addcontentsline{toc}{section}{Appendices}
\renewcommand{\thesubsection}{\Alph{subsection}}

\subsection{Figures and Tables}
\begin{figure}[ht]
    \centering
    {\includegraphics[width=5cm]{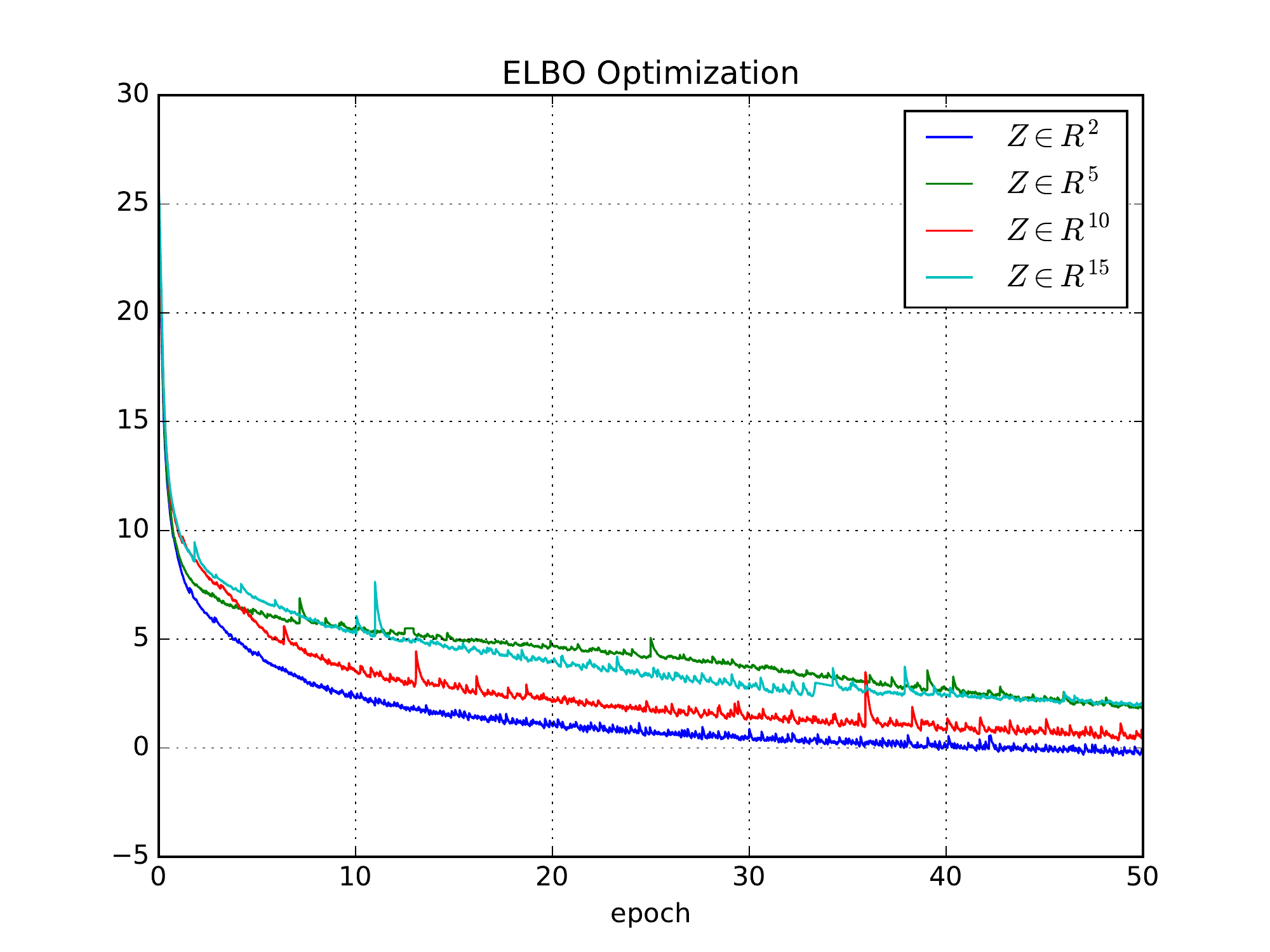}}
	\hspace*{4em}
	{\includegraphics[width=5cm]{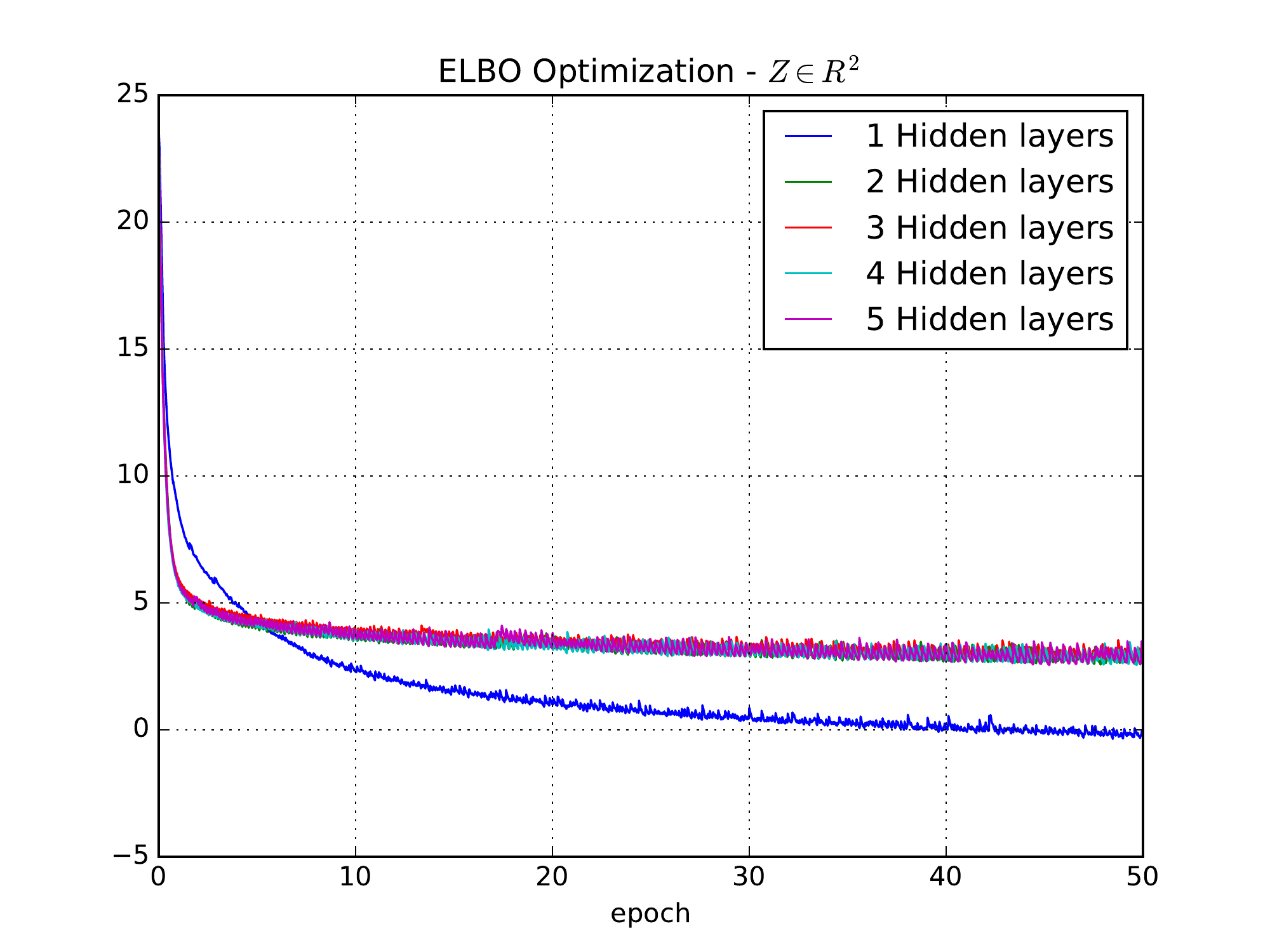}}
	\centering
	{\includegraphics[width=5cm]{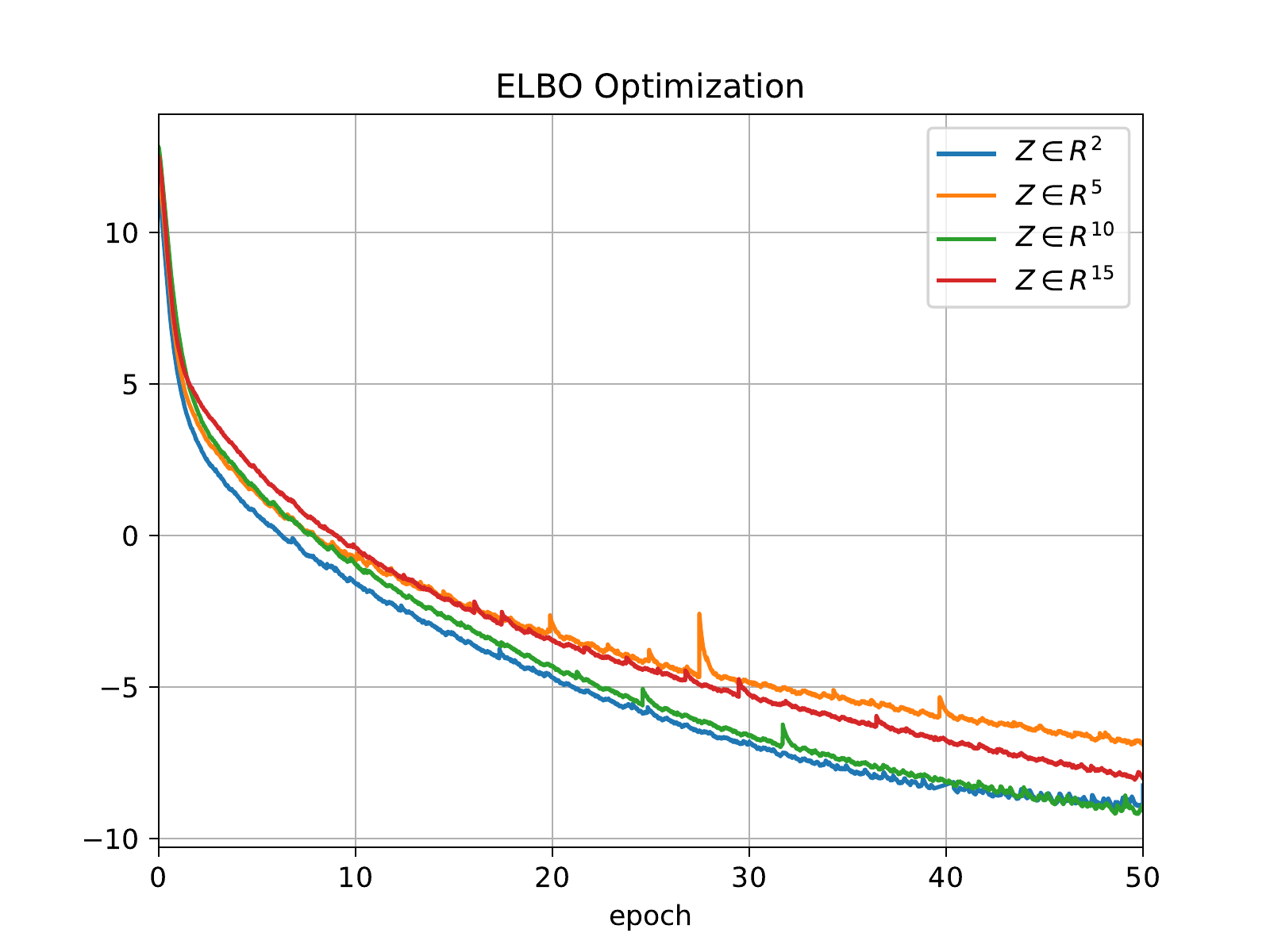}}
	\hspace*{4em}
	{\includegraphics[width=5cm]{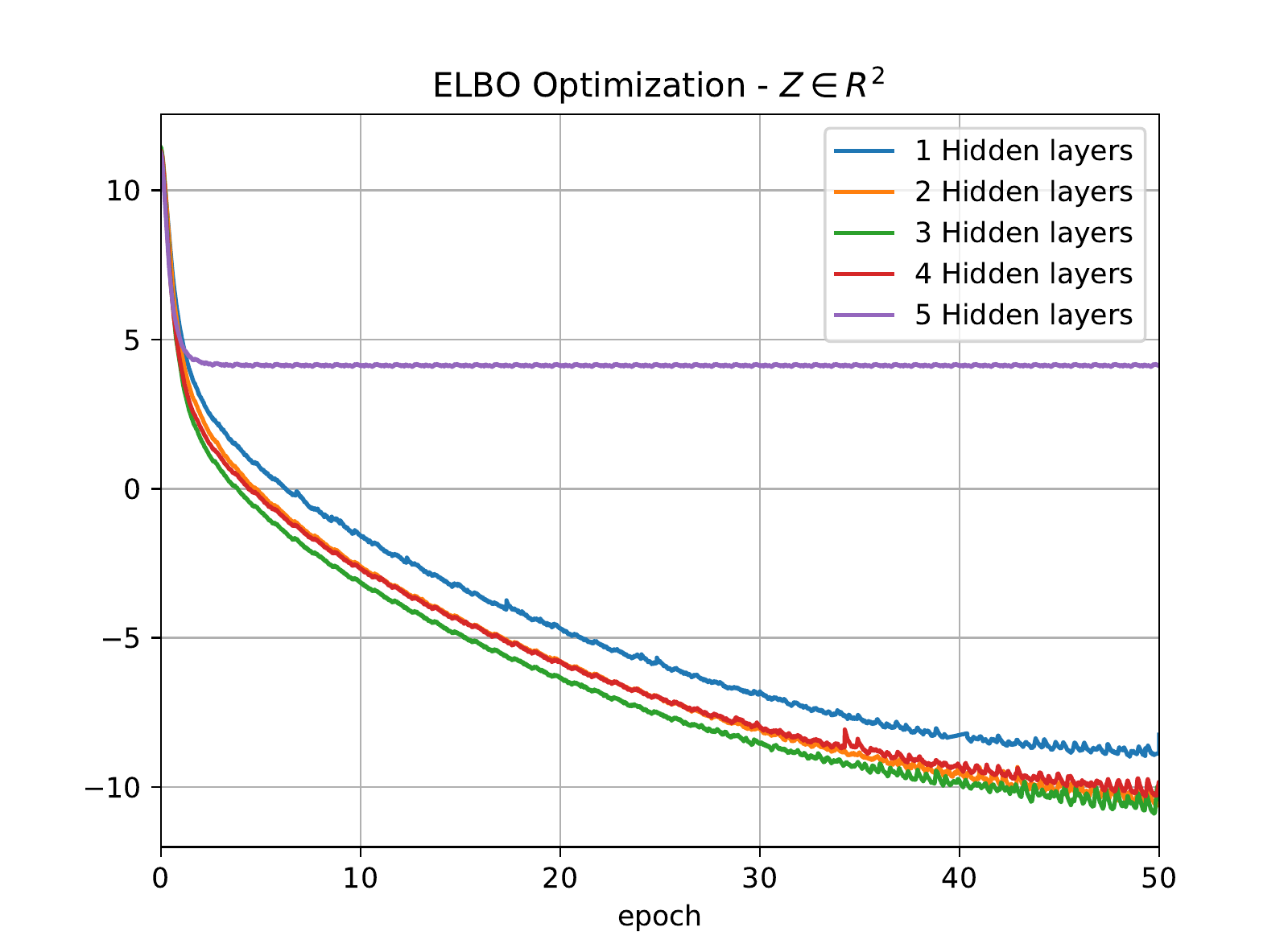}}
	\centering
	{\includegraphics[width=5cm]{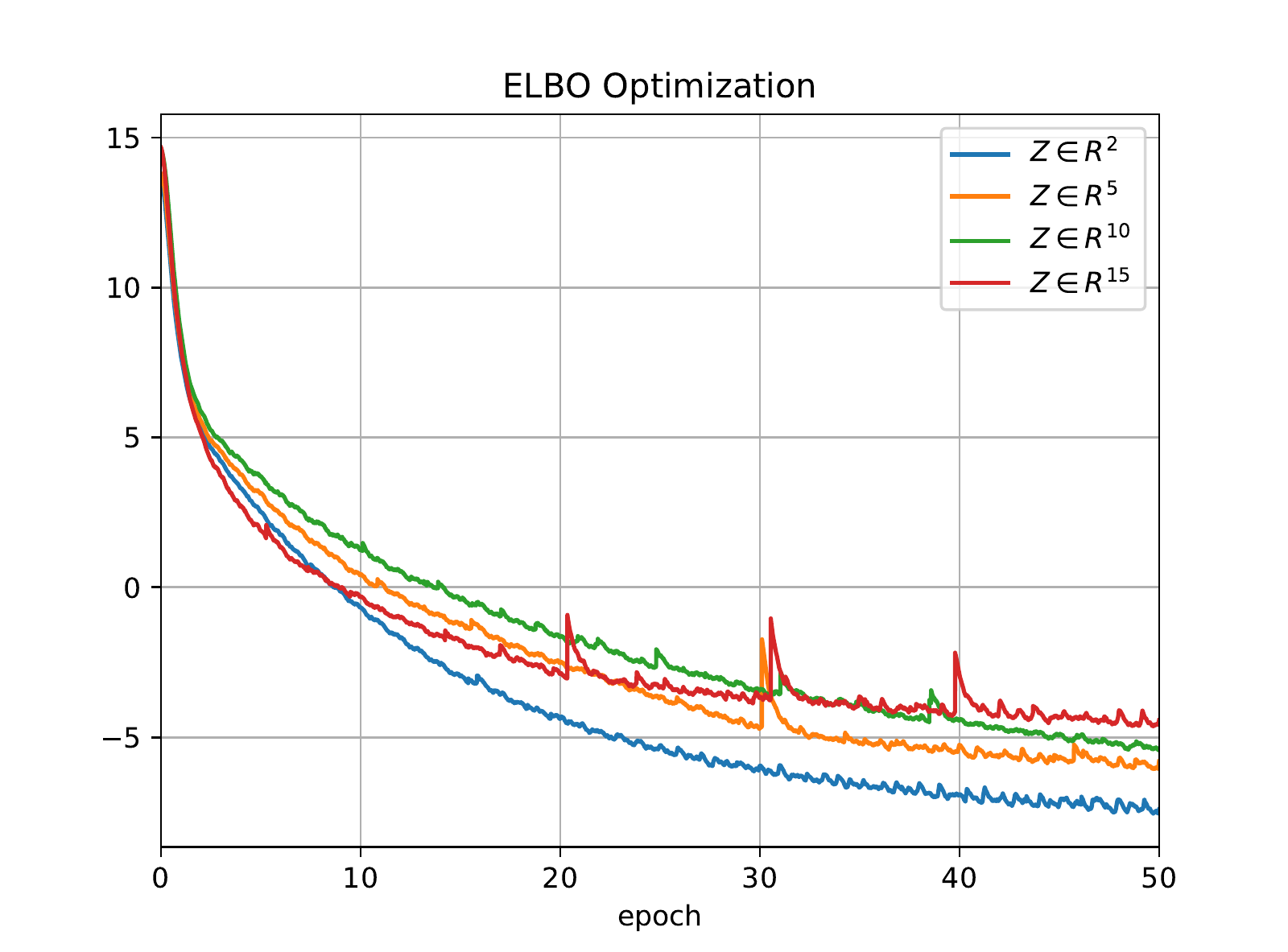}}
	\hspace*{4em}
	{\includegraphics[width=5cm]{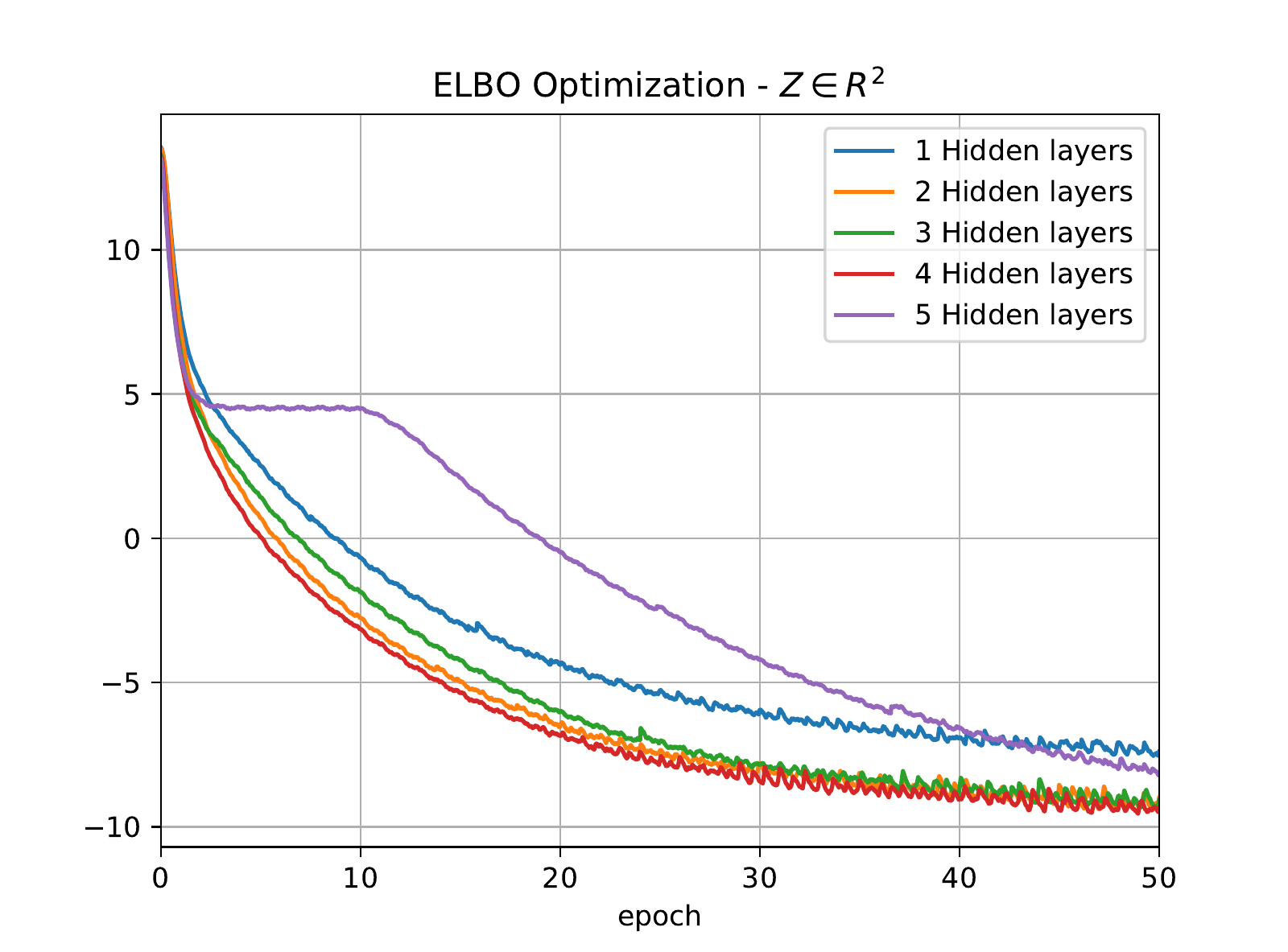}}
	\caption{Panels to the left show the optimization of the negative ELBO for different dimensionalities in the latent space. For $z \in \mathbb{R}^2$, the AEVB algorithm converges faster to the optimal variational density $q^*(z)$ for all datasets (Norwegian dataset top-left panel, Kaggle dataset middle-left and Finnish dataset bottom-left panel). Further, panels to the right also show the optimization of the ELBO but for  $z \in \mathbb{R}^2$ and for a different number of hidden layers. The VAE for the Norwegian dataset (top-right panel) with 1 hidden layer converges faster to $q^*(z)$. For the Kaggle dataset (middle-right panel), 2-4 hidden layers converge faster to the optimal variational density. However, the resulting clustering structure in the latent space contains only two clusters. Similarly, for the Finnish dataset (bottom-right panel) 2-4 hidden layers makes the algorithm converge faster. However, the resulting clustering structure contains four clusters. For this dataset, it is not optimal to have four clusters.}
	\label{elbo_op}
\end{figure}

\begin{figure}[t]
\centering
{\includegraphics[width=5cm]{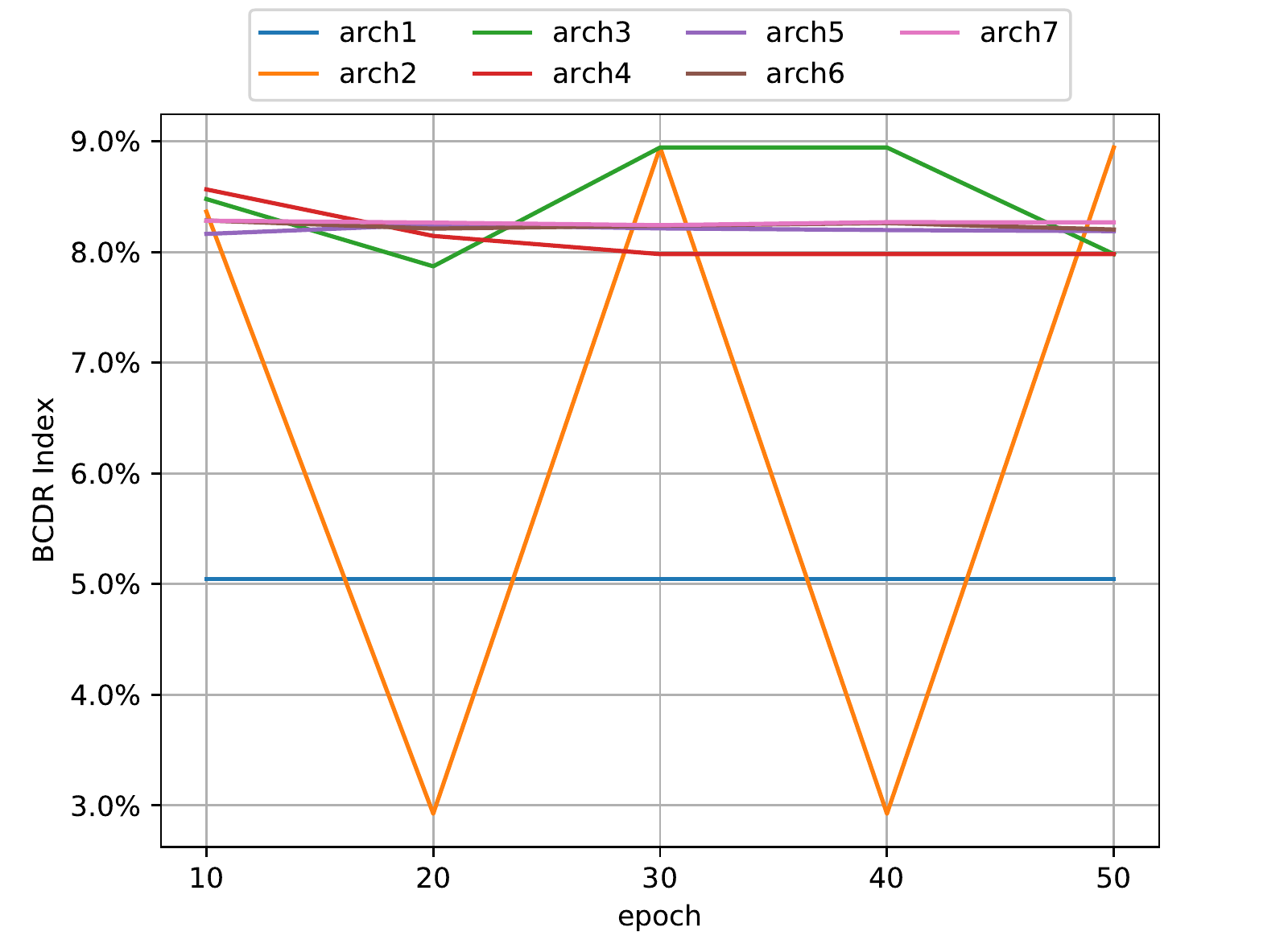}}
\hspace*{4em}
{\includegraphics[width=5cm]{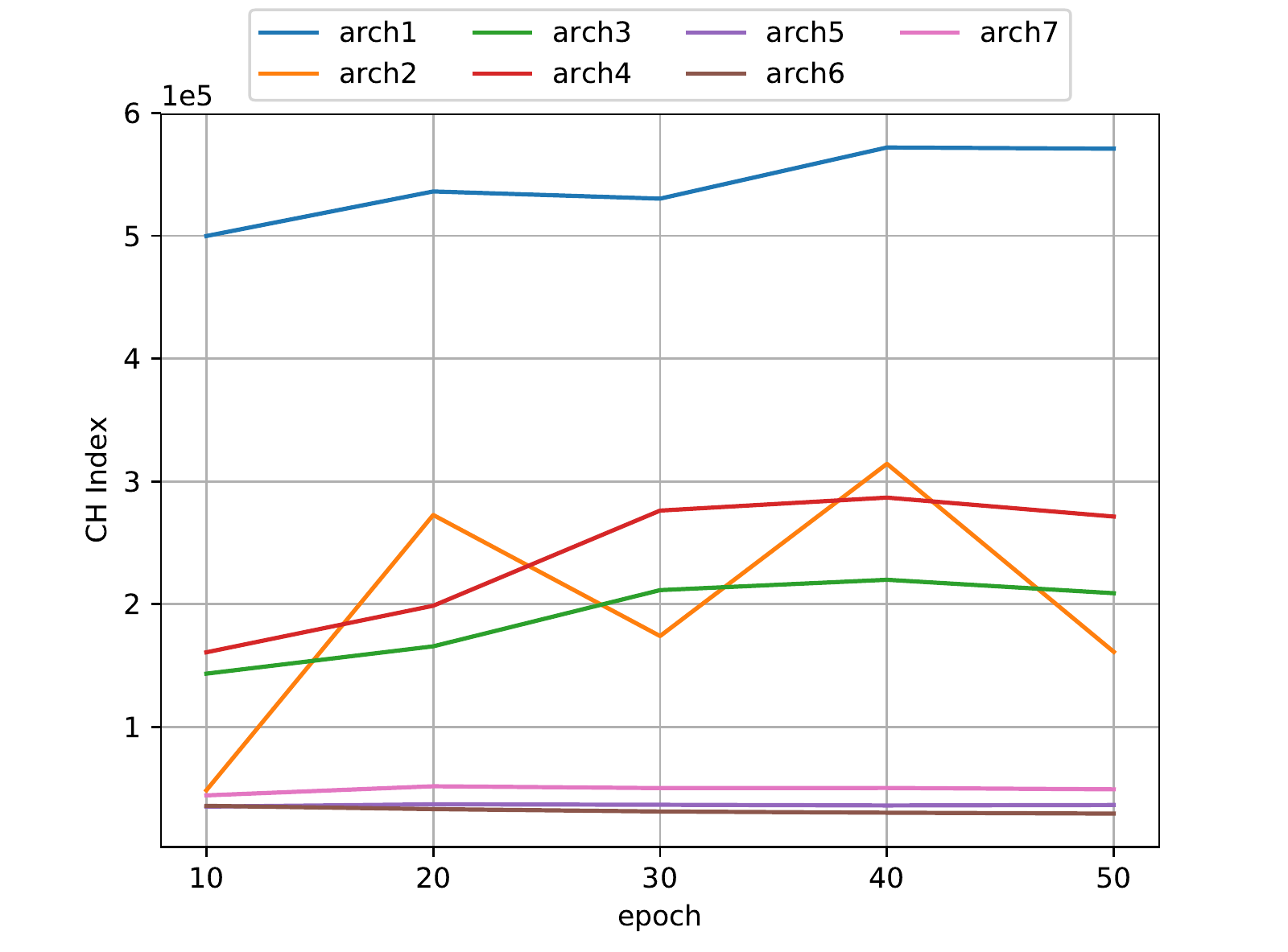}}
\centering
{\includegraphics[width=5cm]{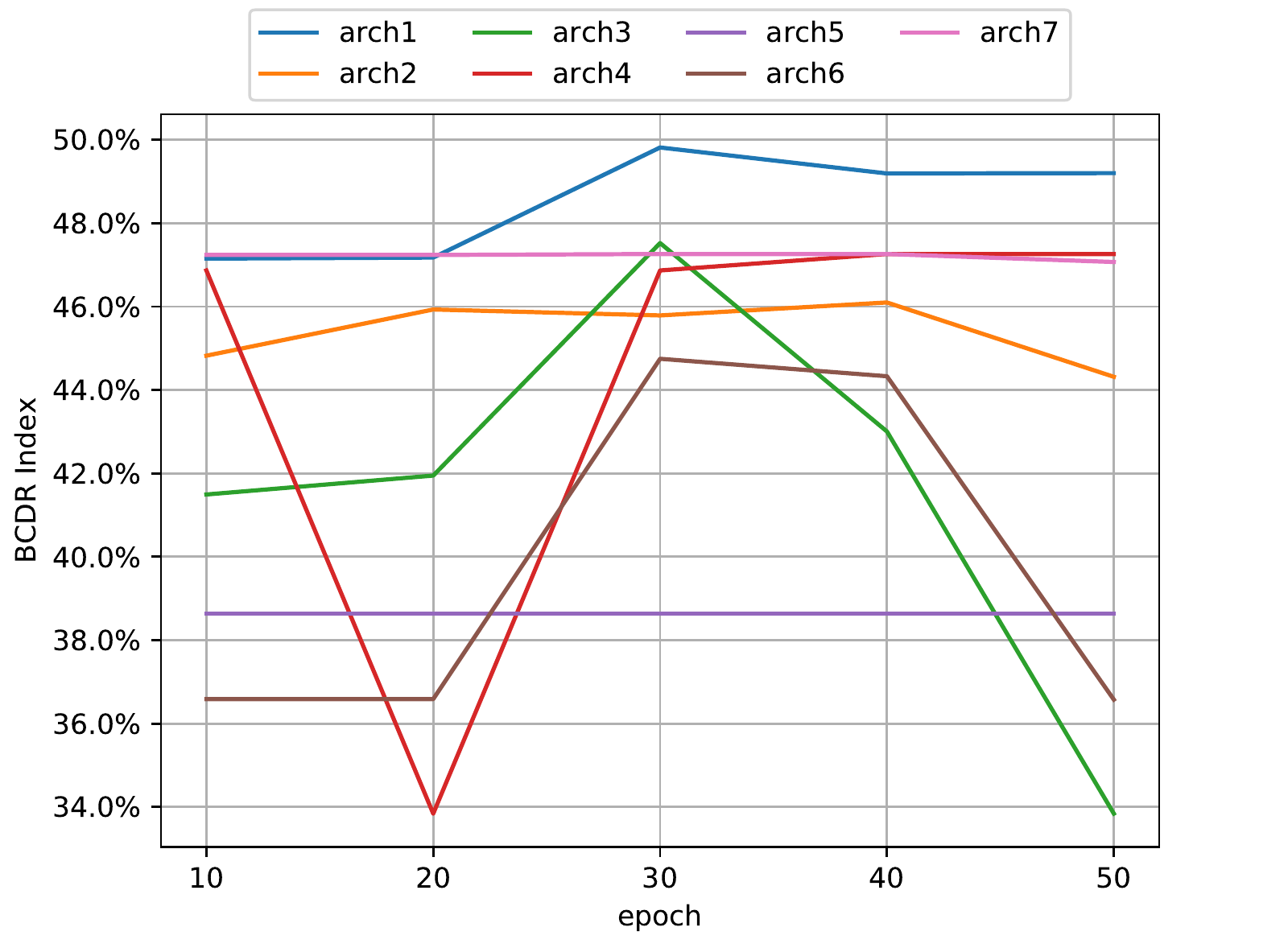}}
\hspace*{4em}
{\includegraphics[width=5cm]{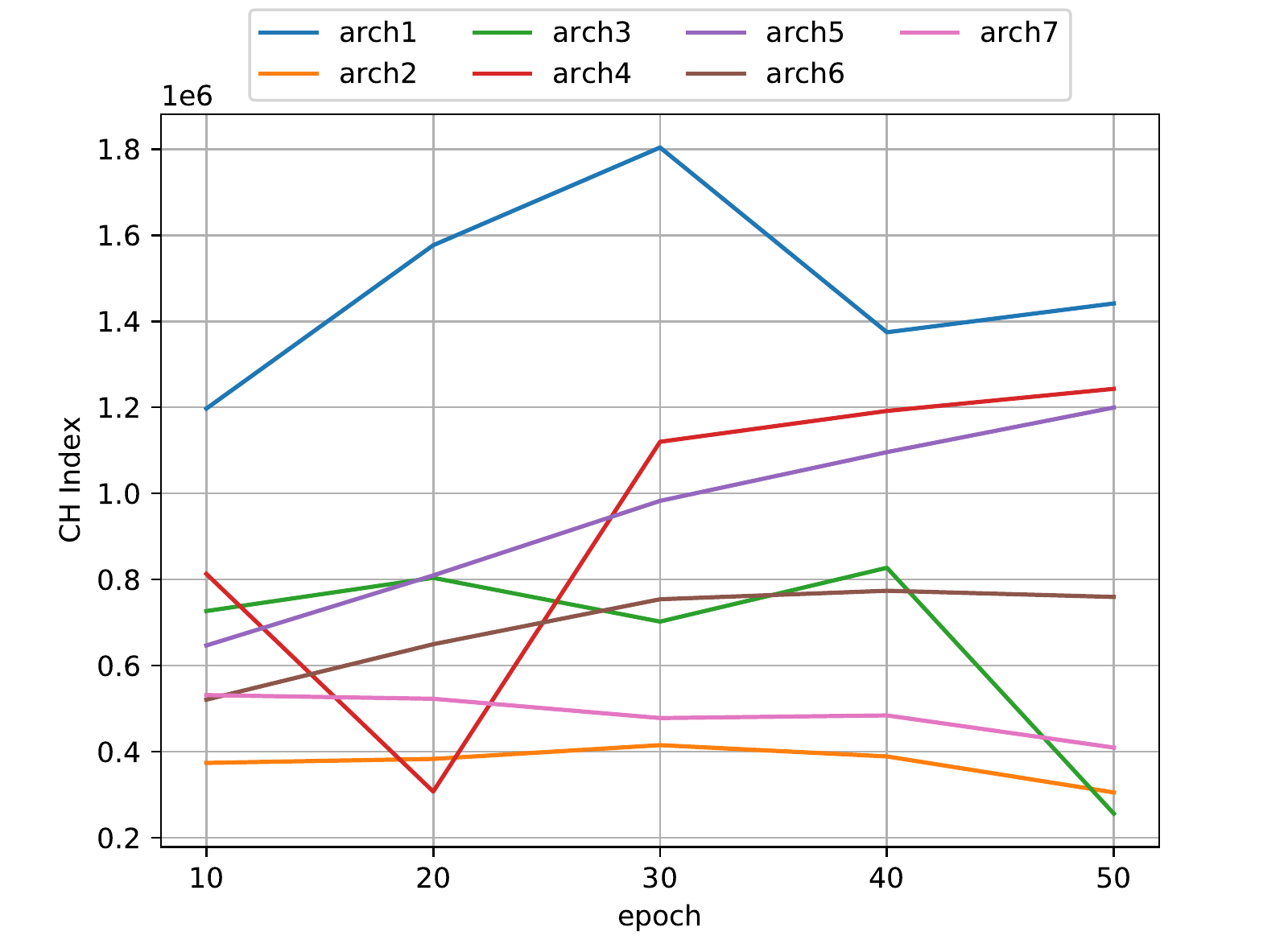}}
\centering
{\includegraphics[width=5cm]{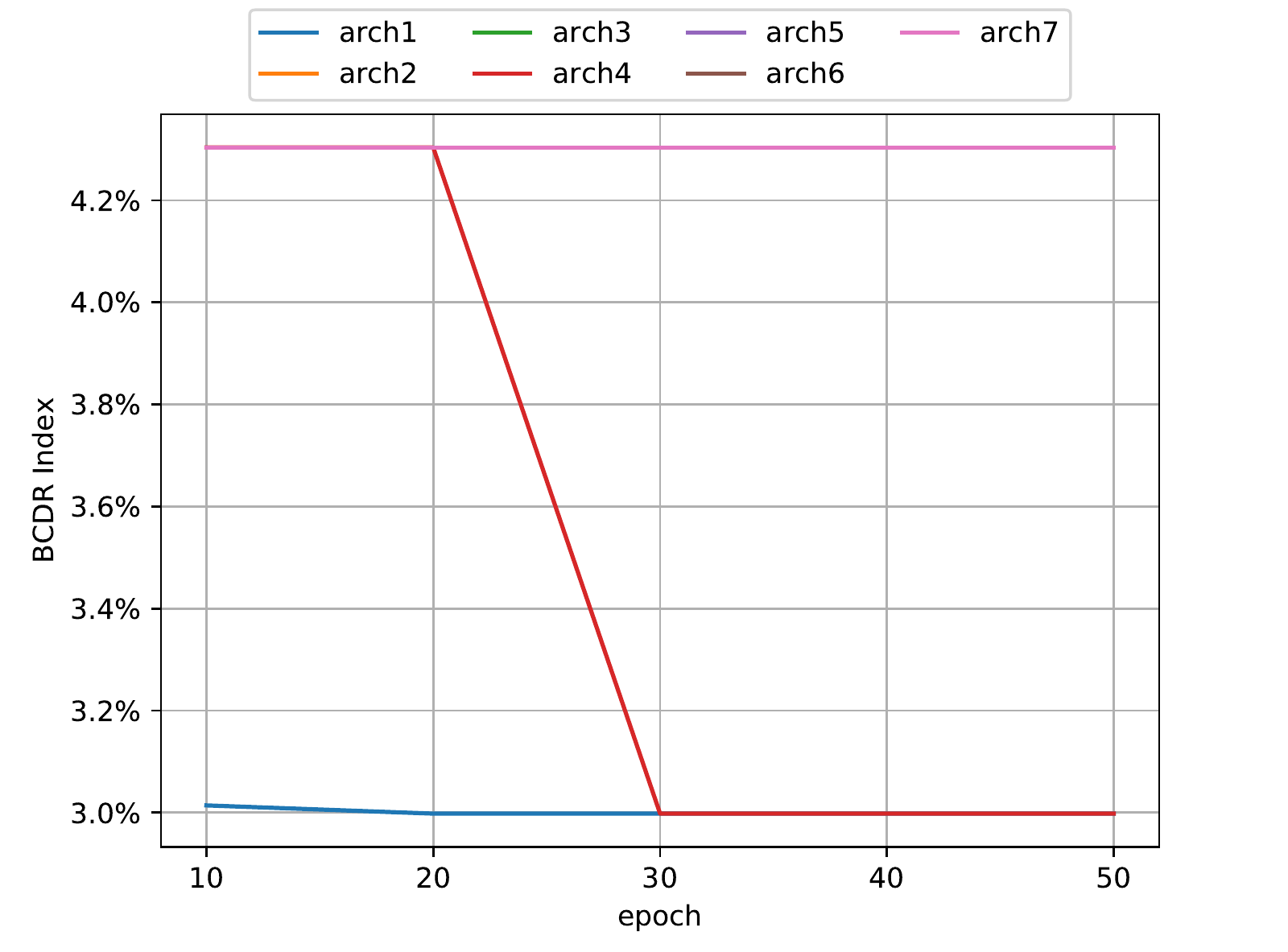}}
\hspace*{4em}
{\includegraphics[width=5cm]{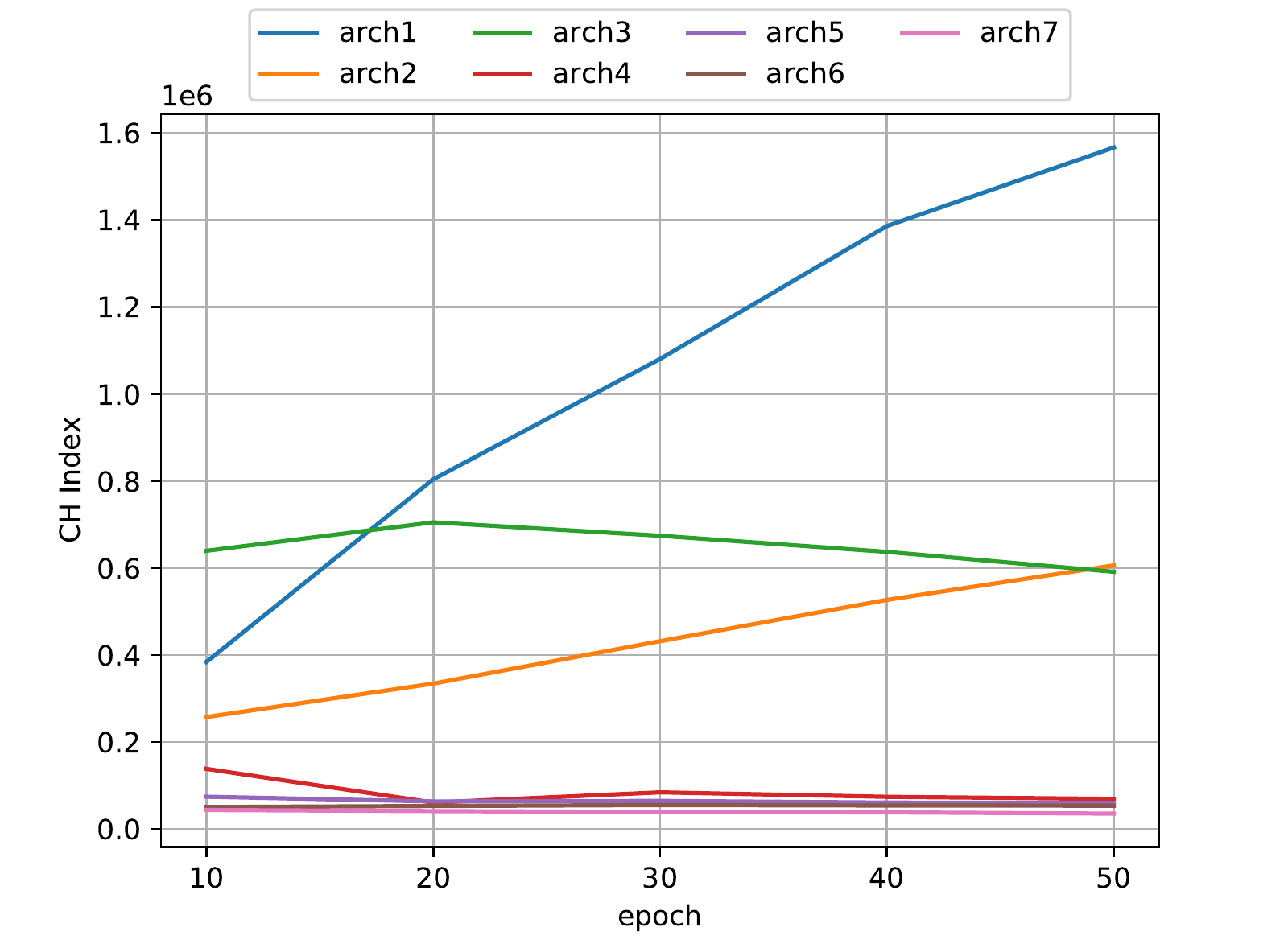}}
\caption{Left panels show the maximum between cluster default rate (BCDR) for the Norwegian (top panel), Kaggle (middle panel) and the Finnish dataset (bottom panel). The panels to the right show the Calinski-Harabaz (CH) index for the same datasets and panels are in the same order. These two cluster metrics, together with the optimization of the ELBO, are used to select the optimal architecture for the VAE. }
		\label{latent_space}
\end{figure}

\begin{table}[ht]
\centering 
\begin{adjustbox}{width=8cm}
\begin{tabular}{ |c|c|c|c|c|c| } 
\hline
Architecture ID & z dimension & Hidden Layers & Neurons & Learning Rate & Epochs \\
\hline
arch1  & 2 & 1 & 5  & 0.01  & 50 \\
arch2  & 2 & 1 & 10 & 0.01  & 50 \\
arch3  & 2 & 1 & 20 & 0.01  & 50 \\
arch4  & 2 & 1 & 30 & 0.01  & 50 \\
arch5  & 2 & 1 & 40 & 0.01  & 50 \\
arch6  & 2 & 1 & 50 & 0.01  & 50 \\
arch7  & 2 & 1 & 60 & 0.01  & 50 \\
arch8  & 2 & 1 & 70 & 0.01  & 50 \\
arch9  & 2 & 1 & 30 & 0.007 & 50 \\
arch10  & 2 & 1 & 30 & 0.008 & 50 \\
arch11 & 2 & 1 & 30 & 0.009 & 50 \\
arch12 & 2 & 1 & 30 & 0.011 & 50 \\
arch13 & 2 & 1 & 30 & 0.012 & 50 \\
arch14 & 2 & 1 & 30 & 0.013 & 50 \\
\hline
\end{tabular}
\end{adjustbox}
\caption{Different architectures tested to train the VAE for the three different datasets. More complex architectures, with more hidden layers and different dimension in the latent spaces, were also tested. However, for the datasets under analysis relative complex architectures do not add any significant value.}
\label{all_archs}
\end{table}

\begin{table}[h]
\centering 
\begin{adjustbox}{width=5.5cm}
\begin{tabular}{ |c|c|c|c| } 
\hline
Name                & Cases   & Features & Default rate \\
\hline
Norwegian car loans  &  187 069  & 20       & 0.0137     \\
Finnish car loans     &  115 899  & 12       & 0.0081  \\ 
Give me some credit  &  150 000  & 10       & 0.0668    \\
\hline
\end{tabular}
\end{adjustbox}
\caption{Summary of the three datasets used in the different experiments in this paper. Default rate is defined as $dr_{C_j} = \frac{\sum_i^n c_i[y_i=1]}{n}$, where $C_j$ refers to the \textit{j'th} dataset, $c_i$ is the \textit{i'th} customer in $C_j$ with ground truth class $y_i$.}
\label{tbl_summary_dta}
\end{table}

\begin{table}[ht]
\centering 
\begin{adjustbox}{width=7cm,totalheight=5cm}
\begin{tabular}{ |c|c| } 
\hline
\multicolumn{2}{|c|}{\textbf{\large{Norwegian car loans}}} \\
\hline
Variable Name           & Description                                  \\
\hline
BureauScoreAge          & Matrix with bureau scores and applicants age  \\
NetincomeStability      & Net income stability index                    \\
RiskBucketHistory       & Delinquency history                           \\
NumApps6M               & Number of applications last 6 months          \\
ObjectGroupCarMake      & Car brand in the application                  \\
DownPaymentAgeObject    & Matrix with down payment and car model year   \\
CarPrice                & Car price                                     \\
NetIncomet0t1           & Change in applicant's net income              \\
MaxBucketSnapshot       & Delinquency at the time of application        \\
MaxMoB12                & Months on books at the time of application    \\
NetIncomeTaxt0          & Ratio between net income and taxes            \\
AgeObject               & Car model year                                \\
AgePrimary              & Age of primary applicant                      \\
BureauScoreUnsec        & Bureau score unsecured                        \\
DownPayment             & Own capital                                   \\
MaxBucket12             & Maximum delinquency in the past 12 months     \\
TaxAmountt0             & Tax amount paid                               \\
BureauScore             & Bureau score generic                          \\
Taxt0t1                 & Change in applicant's taxes                   \\
Netincomet0             & Net income at the time of application         \\
\hline
\end{tabular}
\end{adjustbox}
\caption{Variable name and description of all features in the Norwegian car loan dataset.}
\label{tbl_variables_no}
\end{table}

\begin{table}[ht]
\centering 
\begin{adjustbox}{width=1\textwidth}
\begin{tabular}{ |c|c|c|c| } 
\hline
\multicolumn{2}{|c}{\textbf{\large{Kaggle}}}    & \multicolumn{2}{|c|}{\textbf{\large{Finnish car loans}}}            \\
\hline
Variable Name                           & Description                                   & Variable Name         & Description                                   \\
\hline
RevolvingUtilizationOfUnsecuredLines    & Total balance on credit lines                 & AgePrimary                        & Age of primary applicant                          \\
AgePrimary                              & Age of primary applicant                      & AgeObjectContractTerm             & Matrix with car model year and number of terms    \\
NumberOfTime3059DPD                     & Number of times borrower has been 30-59 dpd   & DownPayment                       & Own capital                                       \\
Monthly debt payments divided by monthly gross income  & MaritalStatus                  & Marital Status                    & DebtRatio                                         \\
Income                                  & Monthly Income                                & MaxBucket24                       & Maximum delinquency in the past 24 months         \\
NumberOfOpenCreditLines                 & Number of loans or credit cards)              & MonthsAtAddress                   & Number of months living at current address        \\
NumberOfTimesDaysLate                   & Number of times borrower has been 90 dpd      & Number2Rem                        & Number of 2nd reminders last year                 \\
NumberRealEstateLoansOrLines            & Number of mortgage loans                      & NumberRejectedApps                & Number of rejected applications                   \\
NumberOfTime6089DPD                     & Number of times borrower has been 60-89 dpd   & ObjectPrice                       & Car price                                         \\
NumberOfDependents                      & Number of dependents in family                & ResidentialStatus                 & Whether the applicant owns a house                \\
                                        &                                               & ObjectMakeUsedNew                 & Matrix with car make and whether it is new or used\\
                                        &                                               & EquityRatio                       & Debt to equitity                                  \\
\hline
\end{tabular}
\end{adjustbox}
\caption{Variable name and description of all features in the Kaggle and Finnish car loan datasets.}
\label{tbl_variables}
\end{table}

\begin{table}[ht]
\centering 
\begin{adjustbox}{width=10cm,totalheight=4cm}
\begin{tabular}{ |c|c|c|c|c|c| } 
\hline
\multicolumn{2}{|c|}{\textbf{\large{Norwegian car loan}}} & \multicolumn{2}{c|}{\textbf{\large{Kaggle}}} & \multicolumn{2}{c|}{\textbf{\large{Finnish car loan}}} \\
\hline
Cluster  & Salient Dimension & Cluster  & Salient Dimension & Cluster  & Salient Dimension \\
\hline
1 &  MaxBucket12    & 1 & NumberOfTime3059DPD                   & 1 & AgePrimary        \\
2 &  NetIncomet0t1  & 1 & NumberOfTimesDaysLate                 & 1 &  Number2Rem       \\ 
2 &  MaxBucket12    & 1 & NumberRealEstateLoansOrLines          & 1 & NumberRejectedApps\\
3 &  AgeObject      & 1 & NumberOfTime6089DPD                   & 2 & Number2Rem        \\ 
3 &  NetIncomet0t1  & 2 & RevolvingUtilizationOfUnsecuredLines  & 2 & NumberRejectedApps\\
3 &  Taxt0t1        & 2 & DebtRatio                             & 3 & DownPayment       \\ 
3 &  DownPayment    & 3 & NumberOfTime3059DPD                   & 3 & ResidentialStatus \\ 
4 &  NumApps6M      & 3 & NumberOfTime6089DPD                   &   &                   \\
4 &  AgeObject      & 3 & NumberOfDependents                    &   &                   \\
4 &  NetIncomet0t1  & 4 & NumberOfTime3059DPD                   &   &                   \\
4 &  Taxt0t1        & 4 & NumberOfTimesDaysLate                 &   &                   \\
4 &  DownPayment    & 4 & NumberOfTime6089DPD                   &   &                   \\
5 &  AgeObject      &   &                                       &   &                   \\
5 &  NetIncomet0t1  &   &                                       &   &                   \\
5 &  DownPayment    &   &                                       &   &                   \\
\hline
\end{tabular}
\end{adjustbox}
\caption{Statistically significant salient dimensions for the Norwegian, Kaggle and Finnish dataset. We use $s.d.=1$ to define salient dimensions.}
\label{tbl_saldim}
\end{table}

\clearpage

\subsection{Existing clustering techniques}\label{clustering_analysis}
This section shows how we can find clusters in a dataset using some of the existing clustering algorithms. Specifically, we test the k-means, the hierarchical clustering and the high density-based spatial clustering of applications with noise (HDBSCAN) algorithms for the Norwegian car loan dataset. We use the WoE transformation introduced in section (\ref{sec_segmentation}) because the dataset contains data points with relatively high values and the algorithms that we are testing tend to isolate these extreme data points into one single cluster. Besides, using the WoE allows a direct comparison between the existing clustering algorithms and the new clustering approach presented in this paper.

\subsubsection{K-means}
The first challenge in the k-means algorithm is to select the $k$ parameter, i.e. the number of clusters. There are some approaches to decide the optimal $k$ parameter. For example, let $k=1,2,...,K$ and fit the algorithm for each $k$ value. Further, evaluate the resulting clusters using a metric for cluster quality, e.g., CalinskiHarabaz (CH), Davies-Bouldin (DB) or Silhouette coefficients (SC), and select the $k$ for the optimal CH, DB or SC index value. The higher CH or SC index, the better clustering structure. On the other hand, the lower DB index, the better clustering structure. Different cluster quality indexes do not necessarily agree on the optimal number of clusters in a given dataset. For the Norwegian car loan dataset the CH and Silhouette indexes select two clusters as the optimal $k$ value, while the DB index selects three clusters.

Remember that the original dataset is in a multidimensional space and that the k-means algorithm is only grouping these vectors around $k$ centroids in the original input space. Therefore, clustering the data with k-means does not offer the possibility to visualize the final clusters. To enable visualization we need to use a dimensionality reduction technique at the top of the k-means algorithm, e.g, we can use principal components analysis (PCA) to reduce the input space into a two-dimensional space keeping track of the k-means clustering labels. Figure (\ref{cluster_plots3}) top panel shows the two-dimensional space for the Norwegian car loan dataset together with the two clusters found by the k-means algorithm. 

\subsubsection{Hierarchical clustering}
The hierarchical clustering, together with the dendrogram, offers a clustering technique which suggests the number of clusters in the dataset. The hierarchical algorithm is based on dissimilarity measures between the input vectors and a linkage function which determines which set of vectors are merged. The common approach in hierarchical clustering is that each vector is a cluster by itself, then these vectors are successively merged until all vectors are grouped in one single cluster.   

We use the ward and complete linkage functions, together with the Euclidean distance, because these two linkage functions do not tend to isolate extreme values in clusters with a relatively small number of vectors. The corresponding dendrograms for the ward and complete functions are shown in Figure (\ref{cluster_plots3}) middle panel.  We can use these dendrograms to select the final number of clusters by merging sets of clusters until the marginal cost of merging two sets is too high. However, this  result depends on the type of distance used and we may obtain different results depending on this choice. 

It is important to mention that the hierarchical clustering algorithm has $O(n^2)$ time complexity and requires $O(n^2)$ memory. Hence, we could only use 45 \% of the data sample compared to the amount of data that we used in other clustering algorithms.

\subsubsection{HDBSCAN}
The HDBSCAN algorithm can find clusters in a dataset and also select the number of clusters. There are some parameters in the algorithm that must be chosen and depending on their values the results can be different. One of the most significant parameters is the minimum cluster size. Using 3000 as input for this parameter, we obtain six different clusters. However, 23\% of the dataset is labeled as noise, and it does not belong to any cluster. On the other hand, using 5000 as the minimum cluster size we obtain three clusters, and 12\% of the data is considered noise. Figure (\ref{cluster_plots3}) bottom panel shows these results.

The previous examples show some of the challenges and limitations that can be faced in the existing clustering techniques. These methods seem to be sensitive to extreme observations and these data points are isolated into small clusters. Further, choosing the number of clusters is a challenging task. Different quantitative approaches can suggest a different number of clusters. One major limitation in the existing methods that we have presented, it is the lack of data visualization, and we need to add a dimensionality reduction technique to enable visualization. Finally, in the specific case of segment-based credit scoring, we need to be able to assign new data points to an existing cluster. For some of the existing clustering techniques, this requirement is not straightforward in their original formulation, e.g., the hierarchical clustering and the HDBSCAN algorithm. 

\renewcommand{\thefigure}{B\arabic{figure}}
\setcounter{figure}{0}

\begin{figure}[h]
    \centering
    {\includegraphics[width=5cm,height=4.1cm]{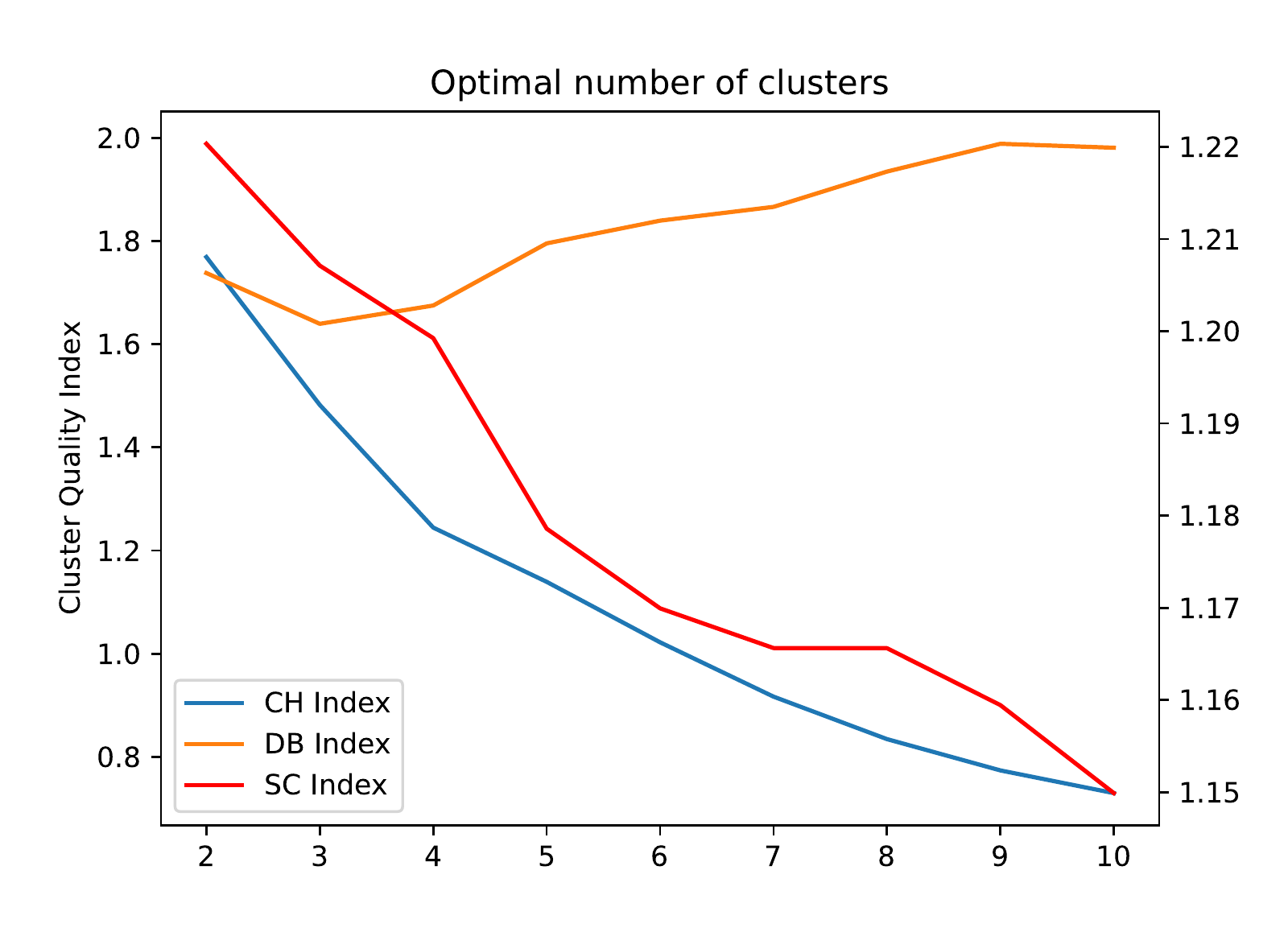}}
	\vspace*{-0.2em}
	{\includegraphics[width=5cm]{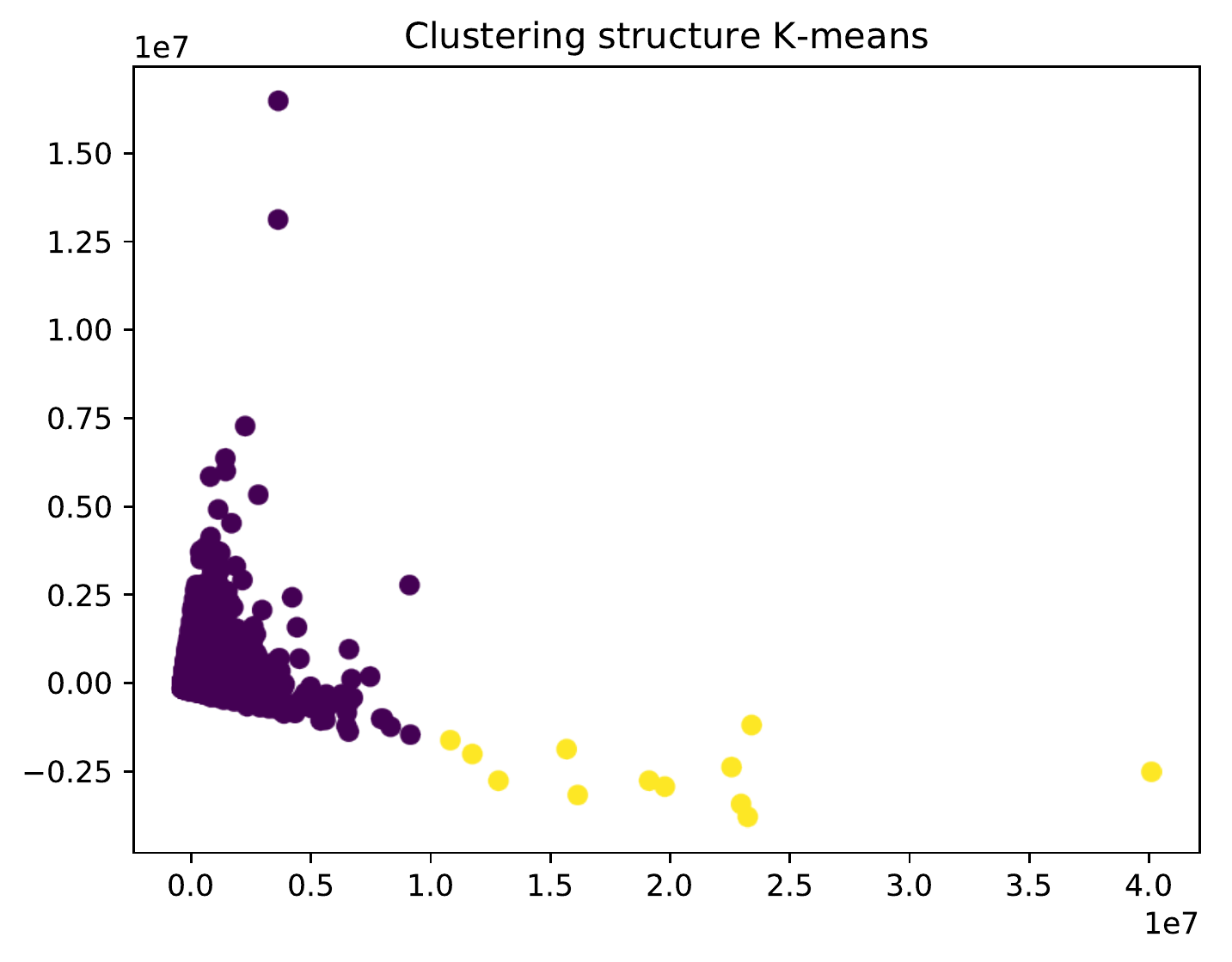}}
	\vspace*{-0.2em}
	{\includegraphics[width=5cm]{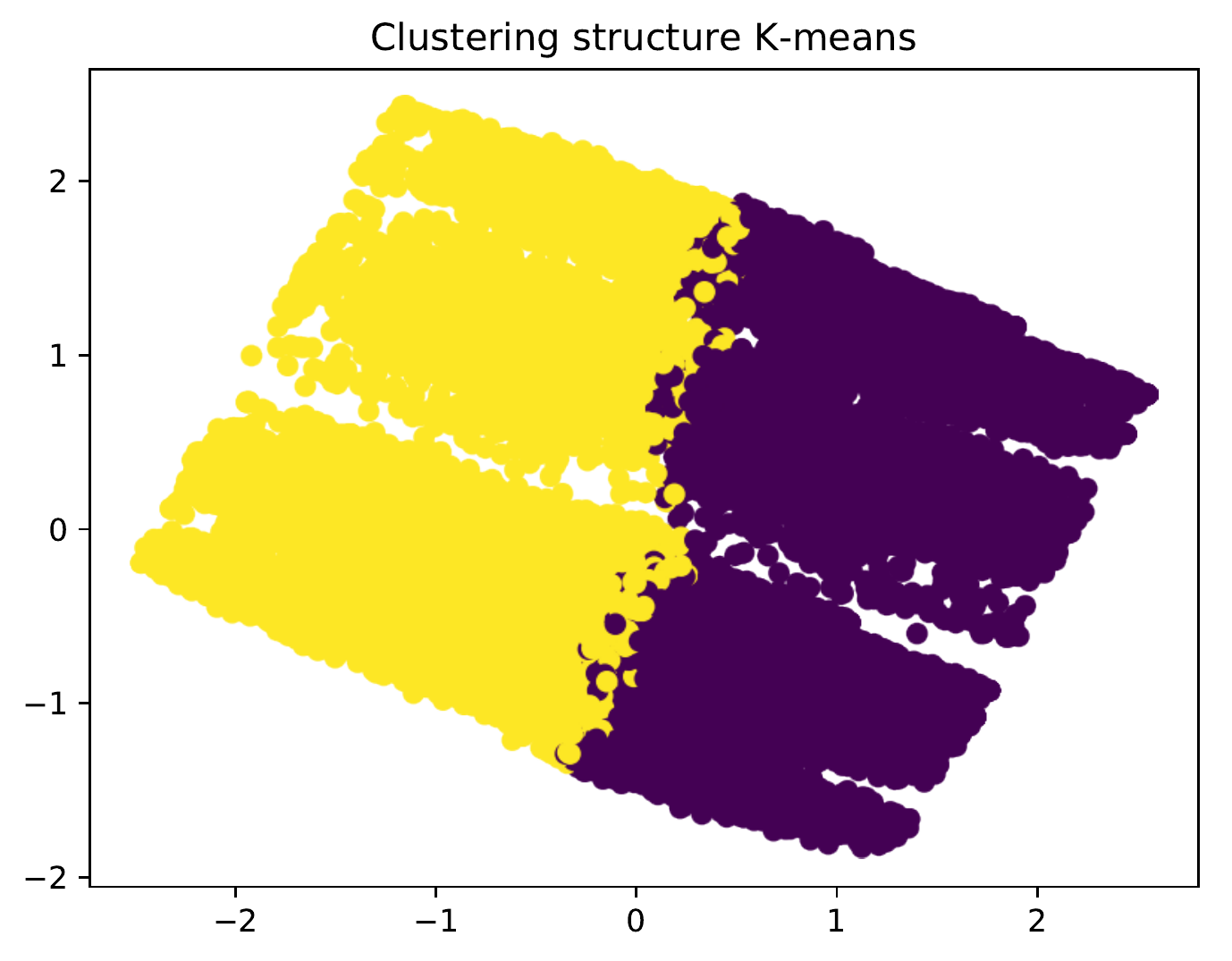}}
	\centering
	{\includegraphics[width=5cm]{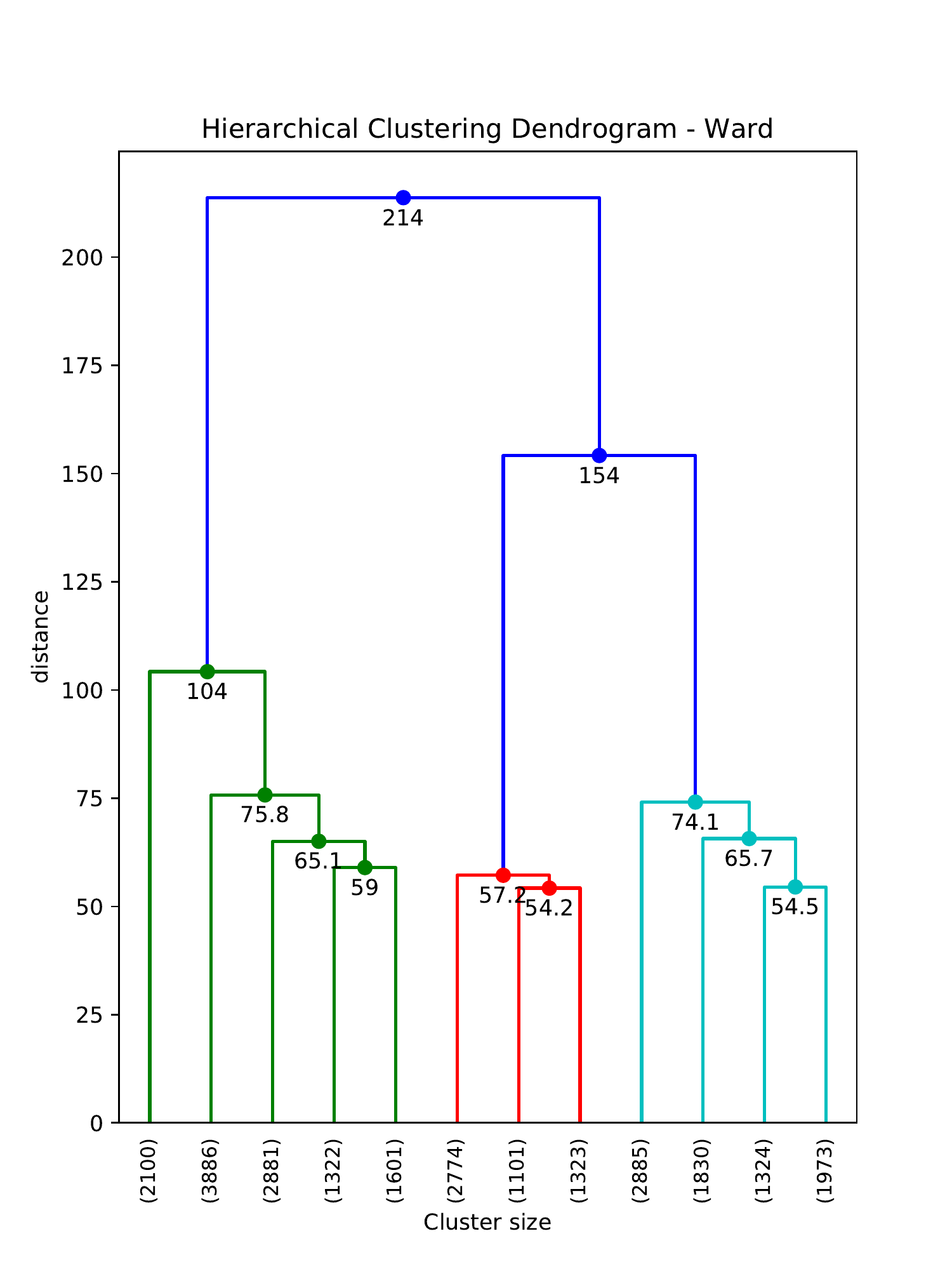}}
	\hspace*{4em}
	{\includegraphics[width=5cm]{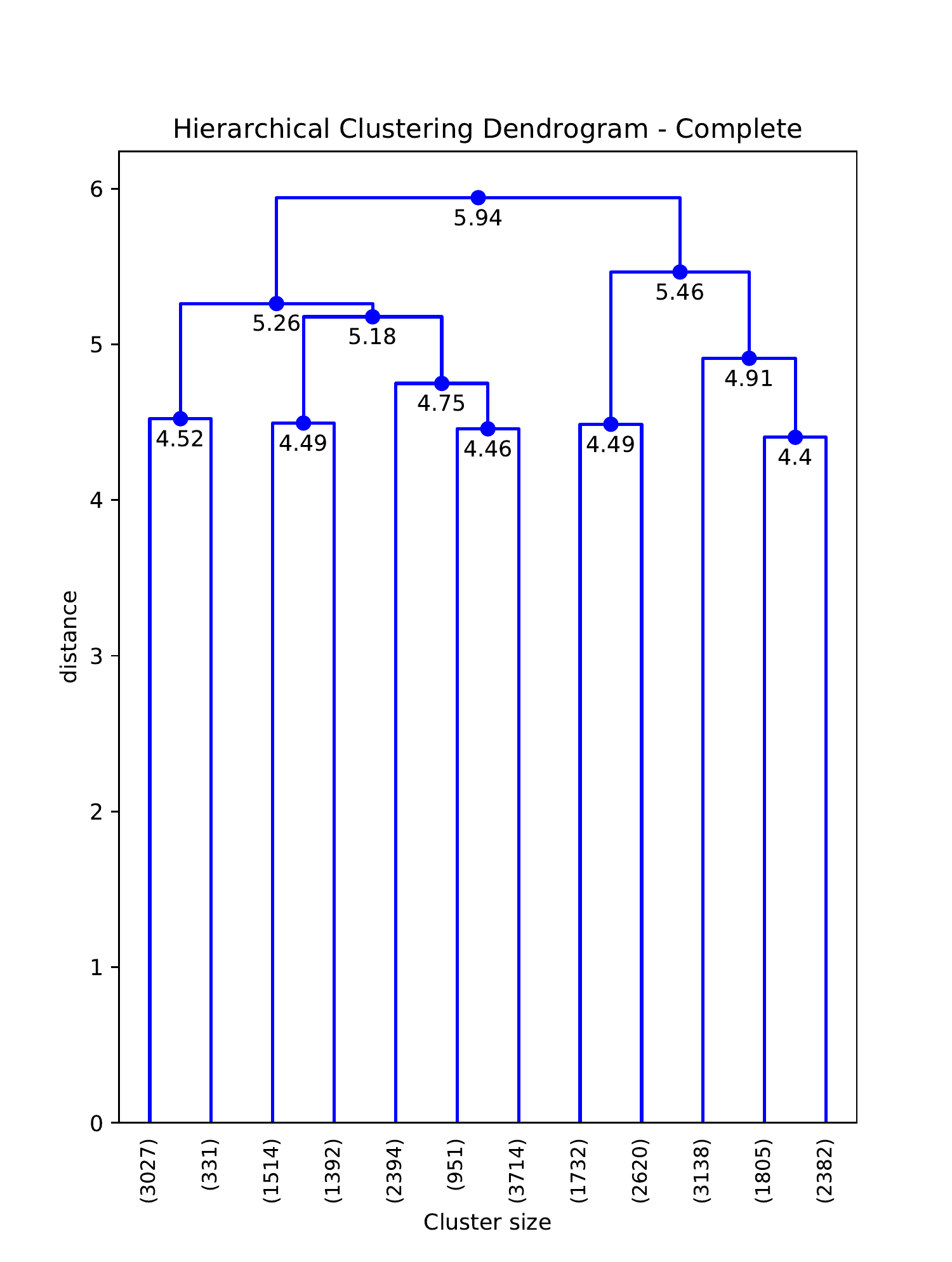}}
	\centering
	{\includegraphics[width=5cm]{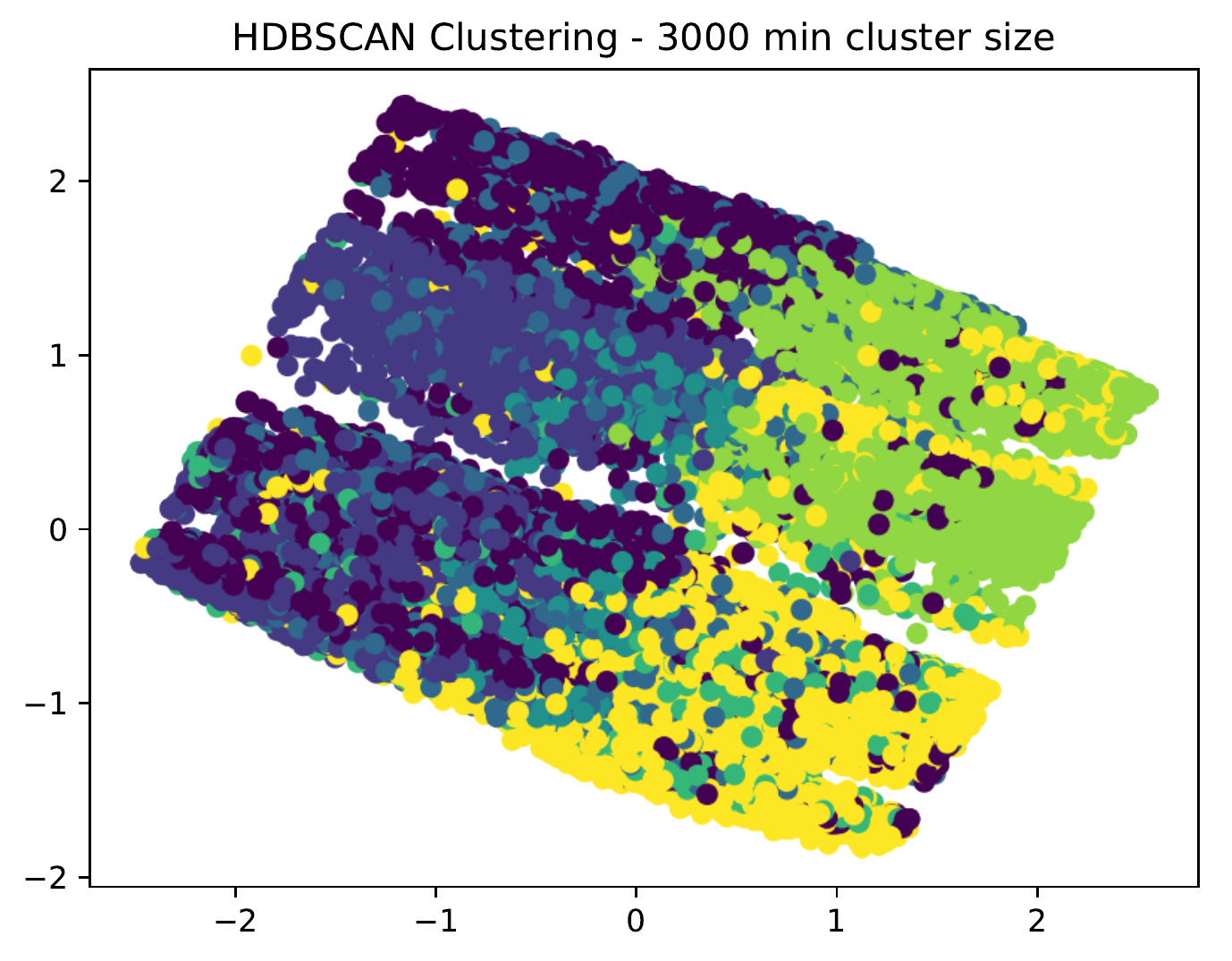}}
	\hspace*{4em}
	{\includegraphics[width=5cm]{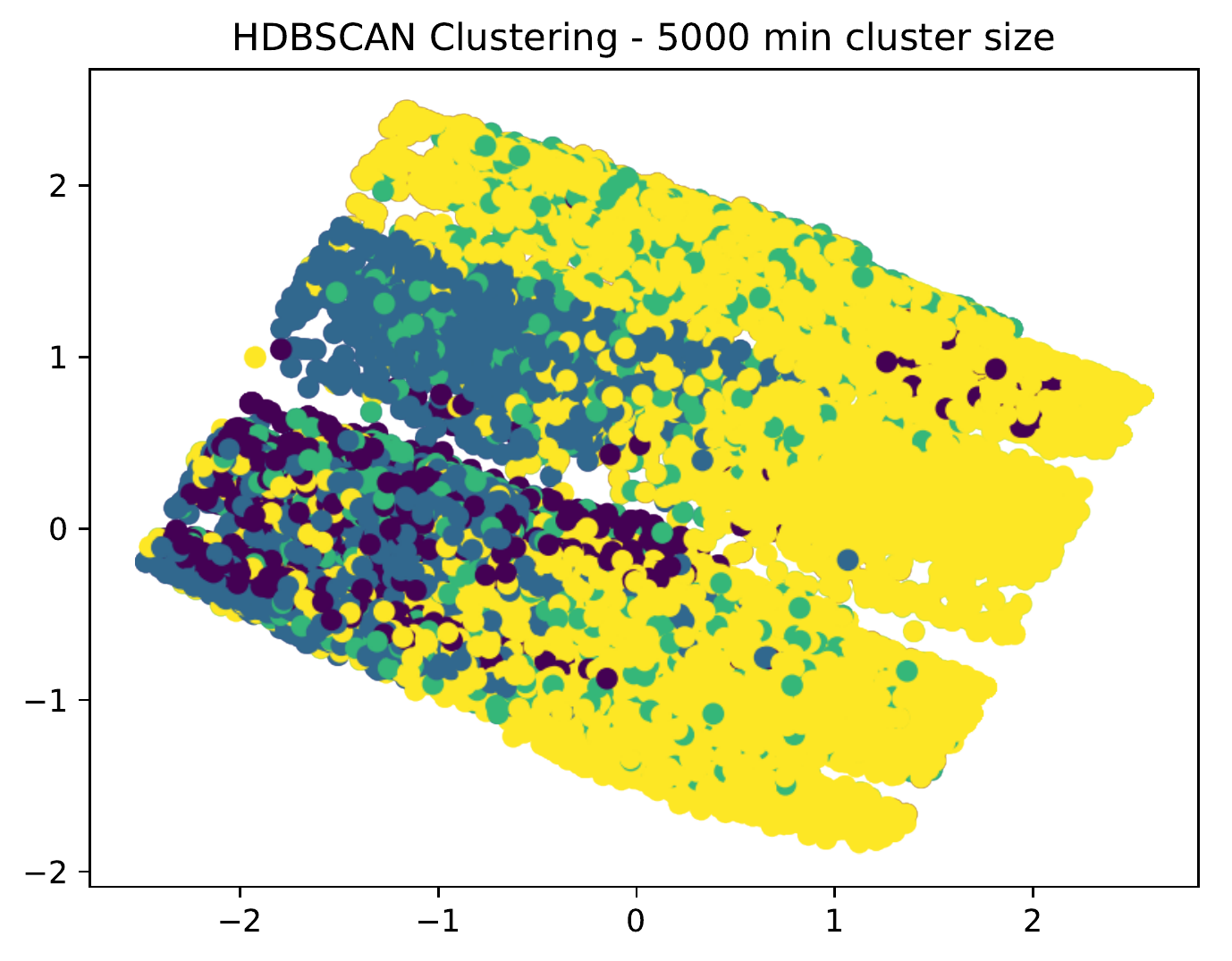}}
	\caption{Top panel: The diagram to the left shows the Calinski-Harabaz (CH), Davies-Bouldin (DB) and Silhouette coefficients (SC) indexes for different $k$ values. Both the figure in the middle and to the right show the first two PCA components together with the k-means labels for $k=2$. The first diagram uses the original dataset, while the second uses the WoE transformation. Middle panel: the first diagram shows the dendrogram for the ward linkage in the hierarchical clustering algorithm, whereas the diagram to the right uses the complete linkage function. Bottom panel: The diagram to the left shows the first two PCA components for the WoE transformation together with the HDBSCAN labels constraining the minimum cluster size to be 3000, while the diagram to the right constraining the minimum cluster size to be 5000. }
	\label{cluster_plots3}
\end{figure}

\end{document}